\documentclass{aa}  

\usepackage{graphicx}
\usepackage{tabularx} % in the preamble
\usepackage{mathtools}
\usepackage{amsmath}  % for the align environment
\usepackage{bm}       % for bold math symbols

\usepackage{mathtools}
\usepackage{blindtext}

\newcommand\numberthis{\addtocounter{equation}{1}\tag{\theequation}}
\DeclareUnicodeCharacter{2212}{-}
\MHInternalSyntaxOn
\renewcommand{\dcases}
 {
  \MT_start_cases:nnnn
    {\quad}
    {$\m@th\displaystyle##$\hfil}
    {$\m@th\displaystyle##$\hfil}
    {\lbrace}
 }
\MHInternalSyntaxOff
\usepackage{natbib,twoopt}
\usepackage[breaklinks=true]{hyperref} %% to avoid \citeads line fills
\bibpunct{(}{)}{;}{a}{}{,}             %% natbib format for A&A and ApJ
\makeatletter
  \newcommandtwoopt{\citeads}[3][][]{\href{http://adsabs.harvard.edu/abs/#3}%
    {\def\hyper@linkstart##1##2{}%
     \let\hyper@linkend\@empty\citealp[#1][#2]{#3}}}
  \newcommandtwoopt{\citepads}[3][][]{\href{http://adsabs.harvard.edu/abs/#3}%
    {\def\hyper@linkstart##1##2{}%
     \let\hyper@linkend\@empty\citep[#1][#2]{#3}}}
  \newcommandtwoopt{\citetads}[3][][]{\href{http://adsabs.harvard.edu/abs/#3}%
    {\def\hyper@linkstart##1##2{}%
     \let\hyper@linkend\@empty\citet[#1][#2]{#3}}}
  \newcommandtwoopt{\citeyearads}[3][][]%
    {\href{http://adsabs.harvard.edu/abs/#3}
    {\def\hyper@linkstart##1##2{}%
     \let\hyper@linkend\@empty\citeyear[#1][#2]{#3}}}
\makeatother
\usepackage{txfonts}
\defcitealias{Gandhi_2023}{G2023}
\defcitealias{Pelletier_2023}{P2023}

\begin{document}

   \title{High-resolution atmospheric retrievals of WASP-76b transmission spectroscopy with ESPRESSO: Monitoring limb asymmetries across multiple transits}

   \author{Cathal Maguire \orcid{0000-0002-9061-780X}\inst{1}\thanks{E-mail:  \href{mailto:maguic10@tcd.ie}{maguic10@tcd.ie}}
          \and
          Neale P. Gibson \orcid{0000-0002-9308-2353}\inst{1}
          \and
          Stevanus K. Nugroho \orcid{0000-0003-4698-6285}\inst{2}\fnmsep\inst{3}
          \and
          Mark Fortune \orcid{0000-0002-8938-9715}\inst{1}
          \and
          Swaetha Ramkumar \orcid{0000-0003-0815-8366}\inst{1}
          \and 
          Siddharth Gandhi \orcid{0000-0001-9552-3709}\inst{4}\fnmsep\inst{5}
          \and 
          Ernst de Mooij \orcid{0000-0001-6391-9266}\inst{6}
          }

   \institute{School of Physics, Trinity College Dublin, University of Dublin, Dublin 2, Ireland
    \and
    Astrobiology Center, NINS, 2-21-1 Osawa, Mitaka, Tokyo 181-8588, Japan
    \and
    National Astronomical Observatory of Japan, NINS, 2-21-1 Osawa, Mitaka, Tokyo 181-8588, Japan
    \and
    Department of Physics, University of Warwick, Coventry CV4 7AL, UK 
    \and
    Centre for Exoplanets and Habitability, University of Warwick, Gibbet Hill Road, Coventry CV4 7AL, UK
    \and
    Astrophysics Research Centre, School of Mathematics and Physics, Queen’s University Belfast, Belfast BT7 1NN, UK
    }

   \date{Received XXXX; accepted XXXX}

% \abstract{}{}{}{}{} 
% 5 {} token are mandatory
 
  \abstract
  {Direct atmospheric retrievals of exoplanets at high-resolution have recently allowed for a more detailed characterisation of their chemistry and dynamics from the ground. By monitoring the longitudinal distribution of species, as well as the varying vertical temperature structure and dynamics between the limbs of WASP-76b, across multiple transits, we aim to enhance our understanding of the 3D nature and chemical/dynamical evolution of such objects over the timescales of months/years. We present retrievals of three VLT/ESPRESSO observations of the ultra-hot Jupiter WASP-76b, including one not yet reported in the literature, from which we constrain the atmospheric abundances, vertical temperature structure, and atmospheric dynamics for both the leading and trailing limbs of the atmosphere separately, via novel rotational broadening kernels. We confirm the presence of VO recently reported in the atmosphere of WASP-76b. We find a uniform longitudinal distribution of Fe and Mg across the limbs of the atmosphere, for each of our transits, consistent with previous works as well as with stellar values. We constrain substellar Na/Fe and Cr/Fe ratios across each of our transits, consistent with previous studies of WASP-76b. Where constrained, V/Fe and VO/Fe ratios were also found to be broadly consistent between the limbs of the atmosphere for each of the transits, as well as with previous studies. However, for two of the transits, both V and VO were unconstrained in the leading limb, suggesting possible depletion due to recombination and condensation. The consistency of our constraints across multiple high-resolution observations, as well as with previous studies using varying modelling/retrieval frameworks and/or instruments, affirms the efficacy of high-resolution ground-based retrievals of exoplanetary atmospheres.}
  % context heading (optional)
  % % {} leave it empty if necessary  
  %  {}
  % % aims heading (mandatory)
  %  {}
  % % methods heading (mandatory)
  %  {}
  % % results heading (mandatory)
  %  {}
  % % conclusions heading (optional), leave it empty if necessary 
  %  {}

\titlerunning{Monitoring limb asymmetries of WASP-76b across multiple transits}
\authorrunning{C. Maguire et al.}
\maketitle

%
%-------------------------------------------------------------------

\section{Introduction}
Ground-based high-resolution (R $\ge 25,000$) observations of exoplanets have recently advanced our ability to constrain the chemical, physical, and dynamical properties of their atmospheres. These observations exploit the presence of thousands of individual planetary absorption and emission lines within a given wavelength range. This sensitivity enables us to unambiguously identify atmospheric species using techniques such as narrowband transmission spectroscopy or the cross-correlation (CC) method \citep{Snellen_2010}. In addition to these powerful methods, advanced techniques at high-resolution, such as the CC-to-logL mapping or direct likelihood evaluation \citep{Brogi_Line_2019,Gibson_2020}, have now allowed us to go beyond simply identifying species to constraining the abundance of atmospheric species, vertical temperature-pressure ($T$-$P$) profile, and dynamical properties via direct atmospheric retrievals both for emission and transmission spectra. \newline \newline  Multiple recent studies have successfully utilised these techniques to constrain the relative abundances of atomic and molecular species, enabling precise constraints on atmospheric C/O ratios and metallicities at high-resolution \citep{Line_2021,Brogi_2023,Boucher_2023,Ramkumar_2023,Smith_2024}. These constraints may offer valuable insights into how these complex objects formed and evolved to their current atmospheric composition and orbital architecture. \newline \newline  These techniques are particularly tailored to close-in giant planets, such as the ultra-hot Jupiters (UHJs). Their extreme proximity to their host stars inevitably ensures a relatively large planet-to-star radius ratio, as well as a short orbital period, which is crucial for the feasibility of observations of transiting systems. Also, under such extreme conditions, the planet's constituent atomic/molecular species undergo thermal dissociation and ionisation \citep{Parmentier_2018} such that heavier refractory species are readily observable as they have not condensed out of the atmosphere. This makes optical high-resolution spectroscopy particularly powerful as these refractory elements possess a large number of absorption/emission lines in this wavelength regime. Strong lateral winds and vertical mixing are also thought to prevent the cold-trapping of condensates in the lower atmosphere of the cooler nightside, circulating them to the hotter dayside where they are vaporised \citep{Mikal_Evans_2022}. Furthermore, as these species have not yet condensed out of the gaseous phase, the observed atmospheric composition and thus metal-enrichment of UHJ may allow us to link these observations directly to theoretical models of planet formation and migration \citep{Oberg_2011,Madhusudhan_2014,Mordasini_2016,Molliere_2022,Lothringer_2021}. In recent years, a plethora of refractory species have been observed in the atmospheres of multiple hot/ultra-hot Jupiters \citep{Hoeijmakers_2018,Nugroho_2020b,Pino_2020}. \newline \newline  These close-in, tidally-locked objects undergo significant rotation during transit, such that we are sensitive to distinct longitudinal regimes of the atmosphere as a function of orbital phase \citep{Wardenier_2021b}. The potential variation in abundance and dynamical properties of these species as a function of orbital phase has opened a new avenue in the characterisation of these objects, by exploring their intrinsic three-dimensional nature. Their tidally-locked orbits also ensure a permanently irradiated dayside, and a cooler nightside, with drastically differing chemical and dynamical properties, which is expected to drive asymmetries in the chemical abundance and $T$-$P$ profile across the limbs of the planet. This effect was first observed in the atmosphere of WASP-76b, in which an apparent ``kink'' in the cross-correlation trail of Fe was observed, hinting at the existence of an asymmetric Fe abundance in the atmosphere across the limbs of the planet, as different regions rotated into and out of view \citep{Ehrenreich_2020}. However, \cite{Savel_2022} have shown that this apparent discrepancy in the Fe abundance across the limbs of the planet is likely explained by a global effect, such as asymmetric cloud formation or a longitudinal temperature asymmetry. This hypothesis gains further support from the presence of a similar ``kink'' in the CC trails of similar neutral atomic species in the atmosphere of WASP-76b \citep{Kesseli_2022, Pelletier_2023}. These species form at similar altitudes as Fe, suggesting that a global process is likely influencing their common apparent radial velocity offsets. \\ \\  Similar asymmetries have also been found in other UHJs. \cite{Prinoth_2022} found absorption signals from a range of neutral and ionised metals in the atmosphere of WASP-189b, with line positions which varied significantly between species, suggesting an inhomogeneity in their spatial distributions. Offsets in velocity between species were also detected in the atmosphere of WASP-121b, particularly for ionised calcium \citep{Maguire_2023}, which is thought to exist as part of an extended outflowing envelope, such that it is present in a different dynamical regime to the aforementioned neutral species. This outflowing envelope of ionised species was further supported by the existence of artificially ``scaled-up" Fe II features relative to Fe, suggesting the model assumption of hydrostatic equilibrium had broken down for Fe II. Further analysis of this data set by \cite{Seidel_2023} found a potential asymmetric super-rotating Na jet, which stems from the planet's hotter dayside post mid-transit. However, as this is only a partial transit of WASP-121b, further analysis is required to rule out its existence as part of a global day-to-night flow.   \\ \\  Recently, this varying longitudinal representation of an exoplanetary atmosphere has been modelled directly at high-resolution with the combination of advanced rotational broadening kernels \citep{Gandhi_2022,Boucher_2023}, and distinct transit lightcurves for the \emph{leading} morning limb and \emph{trailing} evening limb. As a planet transits its host star, the morning limb passes the stellar disk before
the evening limb. For tidally-locked planets, the morning limb is permanently redshifted, whereas the evening limb is permanently blueshifted. This process is, however, further complicated by the existence of complex wind patterns, such as super-rotational equatorial jets or radially outflowing winds, which are often degenerate with planetary rotation. We can use this information to separate the morning and evening contributions of planetary absorption lines in velocity space and separately model each limb. Thus, constraining the atmospheric properties (abundances, $T$-$P$ profiles, dynamics, etc.) of the morning and evening limb separately.\\ \\ \citet[hereafter G2023]{Gandhi_2023} recently utilised this technique to constrain the atmospheric abundance of multiple metals across the limbs of WASP-76b. The abundances were found to be broadly consistent longitudinally, supporting the aforementioned hypothesis present by \cite{Savel_2022} which suggested any asymmetry in a species' CC-trail is likely caused by a global effect, as opposed to a longitudinally varying abundance. \\ \\\citet[hereafter P2023]{Pelletier_2023} recently utilised MAROON-X observations to retrieve a broadly stellar composition of refractory elements for WASP-76b. However, they found substellar abundances relative to Fe for alkali metals, as well as V and Cr, suggesting that a combination of chemical processes is depleting the abundance of several species. They also detect VO in the atmosphere of WASP-76b with both ESPRESSO and MAROON-X observations, which has long been theorised as a thermal inversion agent which drives high-altitude temperature inversions in the upper atmospheres of UHJs, but has until now remained elusive. \\ \\  Similarly to the spatially-resolved analysis of exoplanetary atmospheres at high-resolution, first presented by \cite{Gandhi_2022}, it has been demonstrated that the distinct transit lightcurves caused by asymmetric limbs enable the extraction of separate transmission spectra for morning and evening limbs directly from low-resolution transit time-series measurements \citep{Espinoza_2021,Grant_2022}. These updated modelling approaches will enable the characterisation of the three-dimensional nature of these objects at both low- and high-resolution directly from transmission spectroscopy observations. Coupled with the enhanced sensitivity offered by the next generation space telescopes and ultra-stable ground-based echelle spectrographs, this will also provide a more detailed understanding of the intricate chemical and physical processes at play in their atmospheres.
\\ \\  In Sect. \ref{sec:2} we present our ESPRESSO observations and data reduction, the removal of spectral distortions in the form of the Centre-to-Limb (CLV) variation and Rossiter-McLaughlin (RM) effect, and the removal of stellar and telluric contamination with \textsc{SysRem}. In Sect. \ref{sec:3} we detail our forward model atmosphere, and our rotational broadening kernel, as well as demonstrating our analysis methods. These methods include the cross-correlation technique, as well as outlining our full atmospheric retrieval framework. In Sect. \ref{sec:4} we discuss our findings before concluding the study and summarising our results in Sect. \ref{sec:5}.

% %--------------------------------------------------------------------
\section{ESPRESSO observations}
\label{sec:2}
We observed one full primary transit of WASP-76b with the
ESPRESSO spectrograph \citep{Pepe_2021} at the VLT on 2019 October 18, hereafter referred to as T3. In addition to this observation, we have included two further archival primary transits, observed on 2018 September 2 and 2018 October 30, respectively. These archival transits of WASP-76b were first analysed by \cite{Ehrenreich_2020}, and have been analysed extensively in the subsequent years \citep{Tabernero_2021,Seidel_2021,Gandhi_2022,Kesseli_2022,Gandhi_2023,Pelletier_2023}. They are hereafter referred to as T1 and T2, respectively. Further details regarding these observations are given in Tab. \ref{tab:Table1}. The data were reduced using the ESPRESSO Data Reduction Software (DRS) v3.1.0, and the non-blaze-corrected S2D spectra were extracted.
\begin{table*}
	\centering
	\caption{A summary of the ESPRESSO observations of WASP-76b.}
	\label{tab:Table1}
	\begin{tabularx}{\textwidth}{cccccccc} % four columns, alignment for each
		\hline
		\hline
		Transit & Date & Program ID & Observing Mode & Instrument Mode & N$_{\rm exp}$ (in-transit) & Exp. Time & S/N@550nm \\
		\hline
		\hline
		T1 & 02-09-18 & 1102.C-0744(C) & 1-UT (UT3) & HR21 & 35 (21) & 600 s & $\sim$94 \\
		T2 & 30-10-18 & 1102.C-0744(D) & 1-UT (UT3) & HR21 & 69 (37) & 300 s & $\sim$75 \\
		T3 & 18-10-19 & 0104.C-0642(A) & 1-UT (UT3) & HR11 & 117 (71) & 150 s & $\sim$45 \\
		\hline
	\end{tabularx}
\end{table*}
\subsection{Cleaning and blaze correction}
\label{sec:2_1}
The S2D spectra were then cleaned and blaze corrected following \cite{Merritt_2020} to which we refer the reader. Similarly to \cite{Maguire_2023}, the residual array in our blaze correction was smoothed with a narrower filter than in our previous works (a median filter with a width of 101 pixels and a Gaussian filter with a standard deviation of 200 pixels) in order to minimise the effects caused by the ``wiggle" pattern seen in ESPRESSO spectra \citep{Allart_2020}. To ensure these ``wiggles" --  alongside the relatively aggressive preprocessing steps and stellar and telluric removal procedures -- didn't inadvertently distort our underlying exoplanetary signal, we conducted extensive injection and recovery tests (see Sect. \ref{sec:Inj_Rec} for further details).
\subsection{Removal of CLV and RM effect}
\label{sec:2_2}
In order to optimise the information content of these data sets, it is imperative that we preprocess the data appropriately, removing any distortions caused by stellar and telluric contamination, as well as secondary effects. With transit observations taken with an ultra-stable high-resolution spectrograph such as ESPRESSO, we are also sensitive to distortions in stellar line shapes in-transit caused by the Centre-to-Limb variation (CLV) and Rossiter-McLaughlin (RM) effect. As has become common practice in the reduction of high-resolution transmission spectroscopy observations \citep{Yan_2017,Casasayas_Barris_2017,Casasayas_Barris_2018,Yan_2019,Turner_2020,Nugroho_2020b,Maguire_2023}, we modelled and removed these distortions of the stellar spectral lines, caused by both the CLV and RM effect, from the data, assuming stellar parameters outlined in Tab. \ref{tab:TableA1}. It is worth noting that the efficacy of this procedure relies heavily on a precise estimate of the projected spin-orbit angle, $\lambda$, which has a relatively large uncertainty. A detailed description of this procedure can be found in \cite{Maguire_2023}.
\subsection{Removal of stellar and telluric contamination}
\label{sec:2_3}
In addition to the in-transit distortions caused by the RM and CLV effects, the contamination caused by stellar and telluric lines was also removed. Firstly, we shift each spectrum into the stellar rest frame via the stellar radial velocity, computed via the orbital parameters outlined in Tab. \ref{tab:TableA1}. This ensures that the stellar and telluric lines are (quasi-)stationary in time. These (quasi-)stationary features were then fit with the iterative algorithm \textsc{SysRem}, first applied to high-resolution spectroscopy by \citet{Birkby_2013} and has now become a standard practice in the field \citep[e.g.,][]{Nugroho_2017,Nugroho_2020a,Nugroho_2020b,Birkby_2017,Gibson_2019,Gibson_2020,Gibson_2022,Merritt_2020,Serindag_2021,Maguire_2023,Ramkumar_2023}. Before applying \textsc{SysRem}, each order was divided by its median spectrum, after which the \textsc{SysRem} model is then subtracted from the data. This procedure is akin to fitting the data directly with \textsc{SysRem} and dividing through by the resultant model, however subtraction allows for a faster model filtering procedure whilst retaining the accuracy required for our analysis \citep{Gibson_2022}. This model filtering procedure then in turn replicates any distortions caused by \textsc{SysRem} on our atmospheric model (see Fig. \ref{Broad_model}).\\ \\ In order to determine the ``optimum'' number of \textsc{SysRem} iterations for each data set, an atmospheric model, including all species outlined in Sect. \ref{Forward_model}, was injected with a negative $K_p$ value, where $K_{\rm p}$ is the planet’s radial velocity semi-amplitude, such that it was well separated from the real exoplanet signal. We then performed a likelihood mapping analysis, and the number of iterations which resulted in the optimum detection significance was chosen. In order to compute the detection significance, we followed \cite{Gibson_2020}, and computed a log-likelihood distribution (see Sect. \ref{sec:Retrieval}) as a function of the model scale factor $\alpha$, conditioned on the optimum $\beta$, $K_{\rm p}$, and $\Delta v_{\rm sys}$ values, where $\beta$ and $\Delta v_{\rm sys}$ are the noise scale factor and the systemic velocity offset, respectively. The expected value, or mean, of this distribution was computed, and divided by its standard deviation, resulting in a detection significance. This, in essence, computes the number of standard deviations the maximum signal deviates from an alpha of 0 (i.e., a non-detection). The optimum number of \textsc{SysRem} iterations for each data set was 16, 18, and 22 for T1, T2, and T3, respectively. It is worth noting that this procedure is highly sensitive to the injected model, as a model with many absorption lines in orders dominated by stellar/telluric lines will, in theory, require a higher number of iterations to optimally extract the exoplanet signal. Furthermore, separating each order and subsequently determining the optimum number of \textsc{SysRem} iterations per order would also minimise the inevitable distortion caused by \textsc{SysRem} on the exoplanet's signal \citep{Nugroho_2017}. 
\section{Analysis}
\label{sec:3}
\subsection{Forward model}
\label{Forward_model}
For each of the transits, we performed traditional cross-correlation analysis, as well as a full atmospheric retrieval, with 1D atmospheric models computed using \textsc{irradiator} \citep{Gibson_2022}, including cross-sections from a number of atomic and molecular species (Fe, V, Cr, Mg, Na, and VO). The cross-sections for Fe, V, Cr, and Mg were obtained via the Kurucz line list \citep{Kurucz_1995}. Similarly, the Na cross-sections were obtained via the VALD line list \citep{Piskunov_1995}, whereas the VO cross-sections were obtained via the ExoMol line list \citep{McKemmish_2016}. We defined 70 atmospheric layers, equidistant in log space, ranging from $10^2$ bar to $10^{-12}$ bar, across which we computed a $T$-$P$ profile using the parametric model from \cite{Guillot_2010} with input parameters $T_{\rm irr}$, $T_{\rm int}$, $\kappa_{\rm IR}$, and $\gamma$, corresponding to the irradiation temperature, internal temperature, mean infrared opacity, and the ratio of visible-to-infrared opacity, respectively. After solving for hydrostatic equilibrium, with variable gravity, the opacity of each predefined layer of our atmosphere was computed by weighing each species' cross-section on their respective volume mixing ratios (VMRs), $\chi_{\rm species}$, assuming a constant VMR with altitude. In addition, we included a H$_2$ VMR, $\chi_{\rm ray}$, along with H$_2$ opacities from \cite{Dalgarno_1962}, to account for Rayleigh scattering, as well as a further continuum opacity source caused by an opaque cloud deck at a pressure level $P_{\rm cloud}$, below which the transmission spectrum model is truncated. In practice, we fit for the logarithms of strictly positive parameters, with the exception of temperature. Therefore, in total, our 1D transmission spectrum model has 6 $+$ N$_{\rm species}$ input parameters ($T_{\rm irr}$, $T_{\rm int}$, $\log_{10}(\kappa_{\rm IR})$, $\log_{10}(\gamma)$, $\log_{10}(P_{\rm cloud})$, $\log_{10}(\chi_{\rm ray})$, $\log_{10}(\chi_{\rm species})$). 
\subsection{Broadening kernel}
\begin{figure*}
    \centering
    \includegraphics[width=\textwidth]{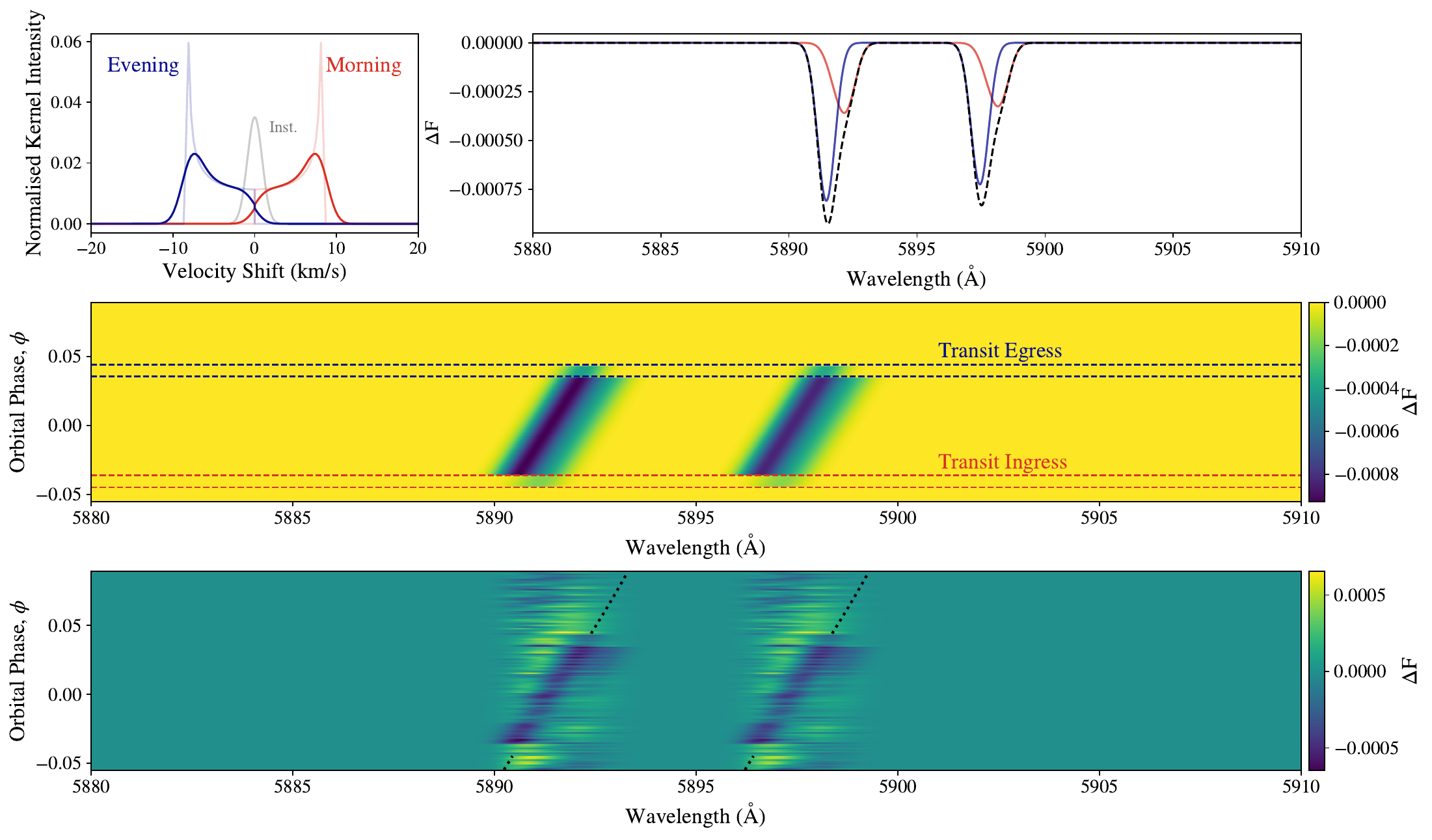}
        \caption{\textit{(Upper Left)} The rotational broadening kernel outlined in Eq. \ref{eqn:kernel}. The solid lines red/blue show the final morning/evening halves, after convolution with an instrumental profile, shown in grey. \textit{(Upper Right)} An asymmetric Na model, with varying broadening and atmospheric parameters between the morning and evening limbs. The combined model is overplotted with a  black dashed line. \textit{(Middle)} The 1D model projected to a 2D orbital phase vs wavelength array for a single ESPRESSO order. This array is then weighted by the limb-darkening model described in \cite{Gandhi_2022}. This ensures solely morning/evening contributions during ingress/egress, respectively. \textit{(Lower)} The 2D orbital phase vs wavelength array after \textsc{SysRem} model filtering, highlighting the distortions of the planetary lines after preprocessing. The black dotted curve shows the planetary velocity to which the model is projected.}
        \label{Broad_model}
\end{figure*}

Recent studies have modelled the varying longitudinal representation of an exoplanetary atmosphere directly with the use of realistic rotational broadening kernels \citep{Gandhi_2022,Boucher_2023}. At high-resolution, we are not only sensitive to the positions of individual absorption lines, but also the subtle variations in their line shapes caused by winds and rotation \citep{Seidel_2020,Hoeijmakers_2020}. Furthermore, with transmission spectroscopy, we are only sensitive to a thin annulus of the planet's atmosphere, i.e., the terminator. \\ \\ 
For a rotating annulus of inner radius $r_1$ and outer radius $r_2 = r_1 + (d\cdot r_1)$, the broadening profile can be modelled by a kernel of the form:
\begin{align}
\label{eqn:kernel}
G(\Delta v)=
\begin{dcases}
    \hfil\frac{a}{d}             & \text{, for   } r_1 \leq \lvert x \rvert < r_2\\
    \frac{a - \sqrt{1 - (\Delta v/\Delta v_{\text{rot}})^2} }{d} & \text{, for   } \lvert x \rvert < r_1
\end{dcases} 
\end{align}
where $a = \sqrt{(1+d)^2 - (\Delta v/\Delta v_{\text{rot}})^2}$, $d$ is the thickness of the annulus relative to $r_1$, $\Delta v$ is the velocity shift of the line at a position $x$, and $\Delta v_{\text{rot}}$ is the shift due to the equatorial velocity. See Appendix \ref{appendix_a} for a full derivation.
\\ \\
This broadening kernel allows us to effectively separate the morning (leading) half of the terminator from the evening (trailing) half of the terminator in velocity, via subtle variations in line shape. In practice, we modelled each half of the terminator (limb) separately with its own 1D atmospheric model and convolved each transmission spectrum with one half of the above broadening kernel. The effect of this convolution is shown in Fig. \ref{Broad_model}. This procedure allows us to retrieve separate atmospheric parameters, as well as separate rotational velocities $\Delta v_{\text{rot}}$ and thickness $d$, for each atmospheric limb. This results in four parameters for our kernel, two for each limb: $d_m$, $d_e$, $\Delta v_{\text{rot}, m}$, and $\Delta v_{\text{rot}, e}$. The subscripts $m$ and $e$ represent the morning and evening limbs, respectively. \\ \\ 
In practice, we fixed $d_m$ and $d_e$ to a constant value equivalent to 5$H_s$ following \cite{Boucher_2023}, where $H_s$ is the atmospheric scale height. We also ensured each half of the kernel normalises to 0.5, such that each 1D model contributed to half the total absorption. This also ensured that any asymmetry in line/transit depth retrieved from the data is caused solely by physical effects, such as an increase in scale height due to a temperature asymmetry between the morning/evening limbs. This asymmetry, or lack thereof, can then be interpreted via varying physical parameters retrieved for the limbs' individual atmospheric models.\\ \\ 
After convolution with our broadening kernels, the independent atmospheric models were summed and linearly interpolated to the 2D wavelength grid of our data (order $\times$ wavelength). For each order, the 2D forward model was then Doppler
shifted to a planetary velocity given by:
\begin{align}
 \label{eq:planet_vel}
  \bm{v}_{\rm p} &= K_{\rm p}\sin{(2\pi\cdot \bm{\phi} )} + \Delta v_{\rm sys} + \bm{v}_{\rm bary}
\end{align}
where $\bm{\phi}$ and $\bm{v}_{\rm bary}$ are the orbital phase and barycentric velocity, respectively. The wavelength solutions have already been corrected for the barycentric velocity by the ESPRESSO DRS. 
This resulted in a 3D shifted forward model with the same dimensions as our data (time/phase $\times$ order $\times$ wavelength). The middle panel of Fig. \ref{Broad_model} shows a single order of our model after this projection. The forward model was then filtered using the \textsc{SysRem} basis vectors via the procedure outlined in \citet{Gibson_2022}. The result of this model filtering is shown in the bottom panel of Fig. \ref{Broad_model}.
\begin{table}[!htbp]
    \centering
    \caption{The parameters of our atmospheric model. $\mathcal{U}(a, b)$ represents a uniform prior from $a$ to $b$. The various parameters for the morning and evening limbs are shown, with a separate $\log_{10}(\chi_{\rm species})$ for each species.}
    \label{tab:Table2}
    \begin{tabularx}{\columnwidth}{*{3}{>{\centering\arraybackslash}X}} % Two columns, centered
        \hline
        \hline \\[-0.3cm]
        Parameter, $\bm{\theta}$ & Units & Prior Range\\
        \hline
        \hline \\[-0.3cm]
        \multicolumn{2}{l}{\hspace{-0.14cm}\textit{Global:}}\\
        % Add your table content here
        $\alpha$ & - & $1$ (fixed) \\[0.15cm]
        $\beta$ & - & $\mathcal{U}(0.1, 2)$ \\[0.15cm]
        $K_{\rm p}$ & km s$^{-1}$ & $\mathcal{U}(150, 250)$ \\[0.15cm]
        $\Delta v_{\text{sys}}$ & km s$^{-1}$& $\mathcal{U}(-20, 20)$ \\[0.15cm]
        \hline \\[-0.3cm]
        \multicolumn{2}{l}{\hspace{-0.14cm}\textit{Morning:}}\\
        $d_m$ &$R_{\rm p}$& $0.062$ (fixed) \\[0.15cm]
        $\Delta v_{\text{rot}, m}$ &pixels& $\mathcal{U}(0, 50)$ \\[0.15cm]
        $\log_{10}(\kappa_{\rm IR})$ &-& $\mathcal{U}(-4, 4)$\\[0.15cm]
        $\log_{10}(\gamma)$ &-& $\mathcal{U}(-4, 2)$\\[0.15cm]
        $T_{\rm irr}$ &K& $\mathcal{U}(1000, 4500)$\\[0.15cm]
        $T_{\rm int}$ &K& $200$ (fixed)\\[0.15cm]
        $\log_{10}(P_{\rm cloud})$ &bar& $\mathcal{U}(-4, 2)$\\[0.15cm]
        $\log_{10}(\chi_{\rm ray})$ &-& $\mathcal{U}(−0.068, 10)$\\[0.15cm]
        $\log_{10}(\chi_{\rm species})$ &-& $\mathcal{U}(-15, -2.3)$\\[0.15cm]
        \hline \\[-0.3cm]
        \multicolumn{2}{l}{\hspace{-0.14cm}\textit{Evening:}}\\
        $d_e$ &$R_{\rm p}$& $0.062$ (fixed) \\[0.15cm]
        $\Delta v_{\text{rot}, e}$ &pixels& $\mathcal{U}(0, 50)$ \\[0.15cm]
        $\log_{10}(\kappa_{\rm IR})$ &-& $\mathcal{U}(-4, 4)$\\[0.15cm]
        $\log_{10}(\gamma)$ &-& $\mathcal{U}(-4, 2)$\\[0.15cm]
        $T_{\rm irr}$ &K& $\mathcal{U}(1000, 4500)$\\[0.15cm]
        $T_{\rm int}$ &K& $200$ (fixed)\\[0.15cm]
        $\log_{10}(P_{\rm cloud})$ &bar& $\mathcal{U}(-4, 2)$\\[0.15cm]
        $\log_{10}(\chi_{\rm ray})$ &-& $\mathcal{U}(−0.068, 10)$\\[0.15cm]
        $\log_{10}(\chi_{\rm species})$ &-& $\mathcal{U}(-15, -2.3)$\\[0.15cm]
        \hline     
    \end{tabularx}
\end{table}
\subsection{Injection and recovery tests}
\label{sec:Inj_Rec}
Before conducting a full retrieval using our atmospheric model, it is imperative that we conduct injection and recovery tests. These tests ensure that the preprocessing steps applied in Sects. \ref{sec:2_1}-- \ref{sec:2_3} do not inadvertently distort the underlying exoplanetary signal, biasing our retrieved constraints and subsequent interpretations. Furthermore, as we aim to constrain any asymmetric signals stemming from the morning/evening limbs, we must ensure that our retrieval framework is sensitive to subtle variations in line shapes, which are already deep beneath the noise of our observations. Thus, typically we are unable to measure individual line shapes directly in order to observe these variations. For each of our data sets, we injected an asymmetrically broadened atmospheric model into a subset of ESPRESSO orders\footnote{We reduced the number of ESPRESSO orders in our injection tests for increased computational efficiency and speed.}, prior to the cleaning and blaze correction steps outlined in Sect. \ref{sec:2_1}. This injected model included cross-sections from Fe, Na, V, and VO. We then performed our retrieval analysis on the injected data, similar to that outlined in Sect. \ref{sec:Retrieval}, and found the retrieved model parameters, $T$-$P$ profiles, and broadening kernels to be in agreement with those injected (see Figs. \ref{T1_inj_retrieval}-- \ref{T3_inj_retrieval}).
\subsection{Atmospheric retrieval}
\label{sec:Retrieval}
\begin{figure*}
    \centering
    \includegraphics[width=0.8\textwidth]{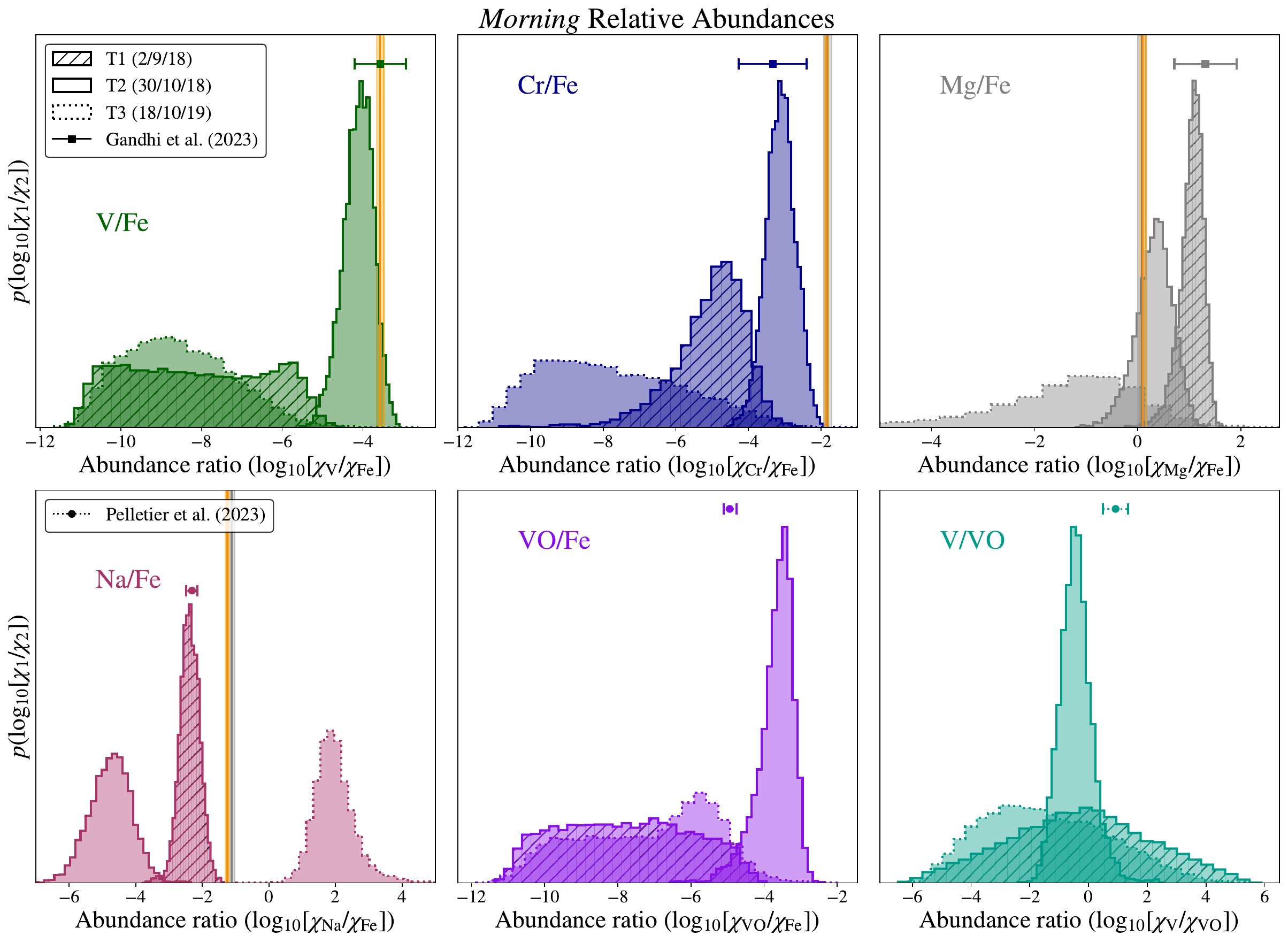}
        \caption{Relative abundance comparison across each of the three transits, for the \textit{morning} limb. The stellar values from \cite{Tabernero_2021} are shown as a grey vertical line with their $1\sigma$ contours shaded, where measured. The solar values from \cite{Asplund_2009} are shown as an orange vertical line with their $1\sigma$ contours shaded. The weighted mean of the retrieved relative abundances in regions C and D (morning) from \citetalias{Gandhi_2023} are also shown, where measured. Otherwise, the relative abundances measured for the whole terminator from \citetalias{Pelletier_2023} are shown. }
        \label{Morn_Relabuns}
\end{figure*}

\begin{figure*}
    \centering
    \includegraphics[width=0.8\textwidth]{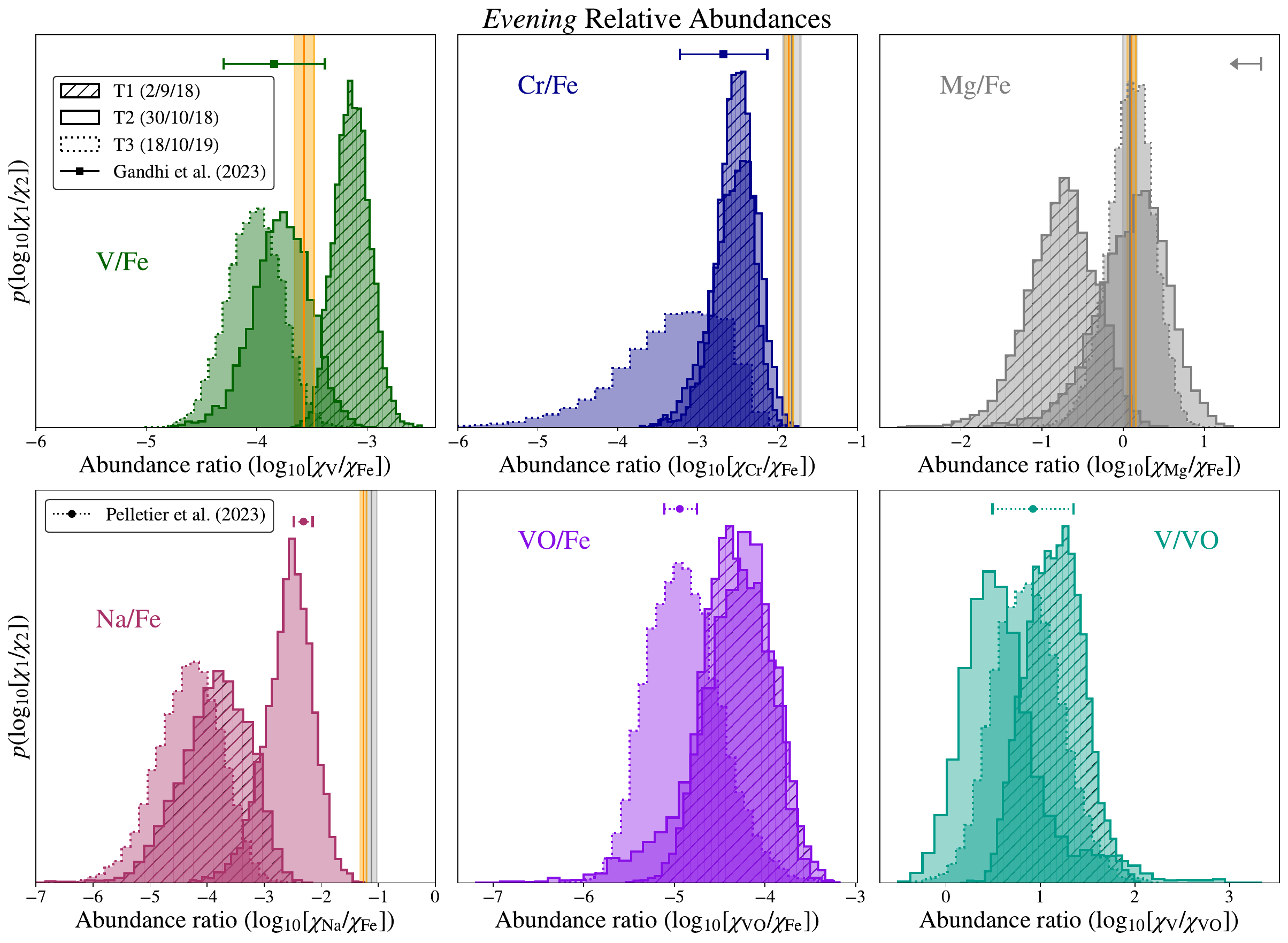}
        \caption{Relative abundance comparison across each of the three transits, for the \textit{evening} limb. Shaded regions are as in Fig. \ref{Morn_Relabuns}. The weighted mean of the retrieved relative abundances in regions A and B (evening) from \citetalias{Gandhi_2023} are also shown, where measured.}
        \label{Eve_Relabuns}
\end{figure*}

In order to perform a full atmospheric retrieval analysis on each of our data sets, we must compute a forward model, $m_i$, for a given set of model parameters, $\bm{\theta}$, outlined in Tab. \ref{tab:Table2}. Assuming uniform priors, the log-posterior is computed by adding the log-prior to the log-likelihood:
\begin{align}
    \ln{\mathcal{L}} = - N\ln{\beta} - \frac{1}{2}\chi^2
    \label{eq:finL}
\end{align}
given
\begin{align}
        \chi^2 = \sum_{i=1}^{N}\frac{(f_i - \alpha m_i(\bm{\theta}))^2}{(\beta\sigma_i)^2}
        \label{eq:chi2}
\end{align}

\noindent where $f_i$ is the data, $\sigma_i^2$ is the data variance, $\alpha$ is the model scale factor, and $\beta$ is the noise scale factor. See \cite{Gibson_2020} for a detailed derivation.\\ \\
This is then folded into an Markov Chain Monte Carlo (MCMC) framework in order to retrieve posterior distributions for each of the model parameters, for each of the data sets. We use a Differential-Evolution Markov Chain (DEMC) \citep{TerBraak_2006,Eastman_2013}, running an MCMC chain with 128 walkers,  with a burn-in length of 300 and a chain length of 600, unless stated otherwise. This results in 115,200 samples of the posterior, of which 38,400 are discarded. The chains' convergence is assessed via the Gelman-Rubin (GR) statistic, after splitting each chain into four independent subchains. We also split the chains into groups of independent walkers, and overplot the 1D and 2D marginal distributions, in order to visualise the convergence of our MCMC. The results of each of our retrievals are shown in Figs. \ref{T1_retrieval}-- \ref{T3_retrieval}. \\ \\ 
 We compare the retrieved \textit{absolute} abundances across each of the transits for the morning (Fig. \ref{Morn_abuns}) and evening (Fig. \ref{Eve_abuns}) limbs. These results are then compared with stellar values, for species with measured stellar abundances \citep{Tabernero_2021}, as well as with solar abundances, where measured \citep{Asplund_2009}.\\\\
Our morning/evening abundance constraints are also compared with the constraints presented by \citetalias{Gandhi_2023}. Due to the relatively large rotation angle of close-in planets such as WASP-76b during transit, \citetalias{Gandhi_2023} opt to retrieve varying regions of the morning/evening limbs before and after mid-transit (see \cite{Gandhi_2022} for further details). We compare our simultaneous retrieval of both limbs with the weighted mean of regions A and B (evening), and C and D (morning) from \citetalias{Gandhi_2023}, respectively. It is also worth noting that the constraints obtained by \citetalias{Gandhi_2023} are from a joint retrieval of T1 and T2, whereas we opt to fit both data sets individually. \\ \\ 
Similarly to previous studies, we also compare the \textit{relative} abundances of our atmospheric species, and find these values to be better constrained (see Figs. \ref{Morn_Relabuns} \& \ref{Eve_Relabuns}). This is due to degeneracies in model transmission spectra, as well as the normalisation of high-resolution spectra during preprocessing, such that we are no longer sensitive to the true continuum of the planet \citep{Birkby_2018}.

\subsection{Cross-correlation}
\begin{figure*}
    \centering
    \includegraphics[width=\textwidth]{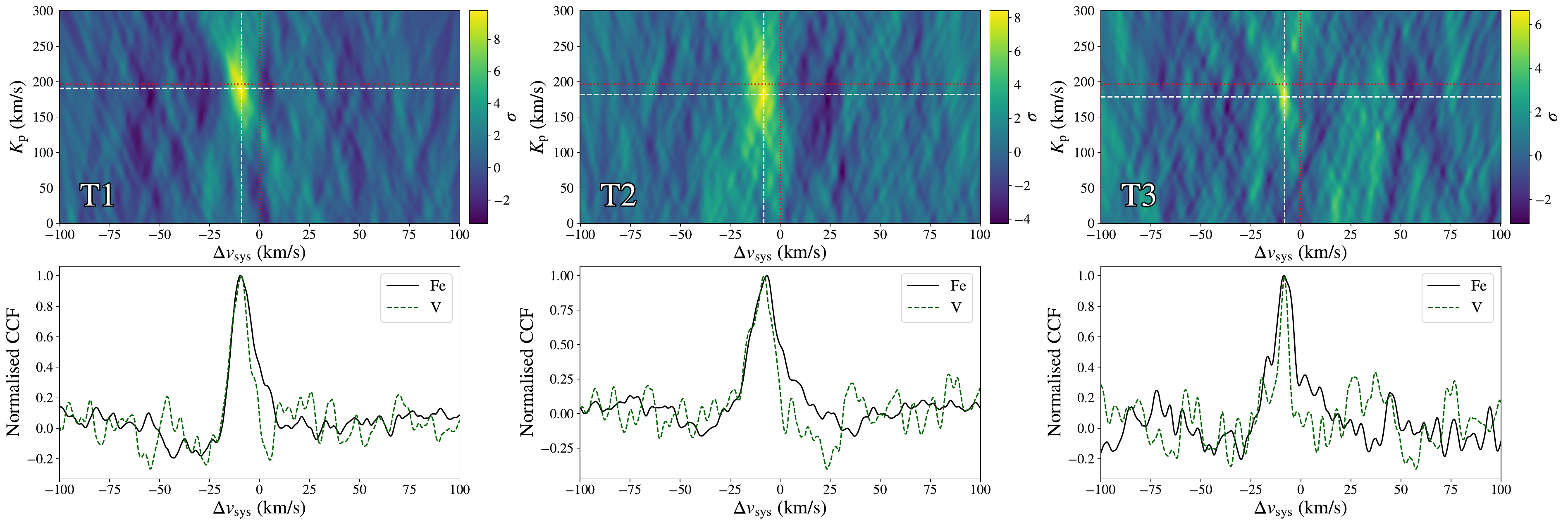}
        \caption{(\textit{Upper}) $K_{\rm p}-\Delta v_{\text{sys}}$ maps of V, for T1, T2, and T3. The V atmospheric models were calculated using the optimum parameters from our various atmospheric retrievals, however, they were unbroadened. The red dotted lines shows the expected planetary velocity values, whereas the white dashed lines shows the maximum $K_{\rm p}$ and $\Delta v_{\text{sys}}$ values. (\textit{Lower}) A comparison of the CCF of Fe and V, at their optimum $K_{\rm p}$ value. As the models were unbroadened, this distribution may be used as a proxy for the average line shape of each species in our data, highlighting any apparent asymmetries in the species' longitudinal distributions.}
        \label{Fe_V_comp}
\end{figure*}

We confirmed the presence of the species included in our atmospheric model individually via traditional cross-correlation analysis. The atmospheric models were calculated using the optimum parameters from our various atmospheric retrievals. The results of which can be seen in Appendix \ref{appendix_d}. We confirm the novel detection of VO in the atmosphere of WASP-76b by \citetalias{Pelletier_2023} in data sets T2 and T3, independently detecting VO in T3 (see Fig. \ref{VO_comp}). As this analysis is the first analysis of the T3 data set, we also aimed to investigate the apparent ``kink'' in the cross-correlation trail first shown by \cite{Ehrenreich_2020} for T1 and T2 (see Fig. \ref{CC_trails}). This apparent ``kink'' has previously been independently confirmed via reanalysis of these transits \citep{Kesseli_2022}, as well as additional analysis from subsequent observations using HARPS \citep{Kesseli_2021} and MAROON-X \citepalias{Pelletier_2023}.
\\ \\ 
Traditional cross-correlation analysis also allows us to investigate any apparent asymmetry in the abundance of individual species between the morning and evening limbs. We first cross-correlate our data with an \textit{unbroadened} model, Doppler shifted over a range of $\Delta v_{\text{sys}}$ values:
\begin{align}
\rm{CCF} (\Delta v_{\rm sys}) &= \sum_{i=1}^{N} \frac{f_i m_i (\Delta v_{\rm sys})}{\sigma_i^2}
 \label{eq:CC}
\end{align}
The cross-correlation function (CCF) for each orbital phase is then shifted to a new planetary velocity given by Eq. \ref{eq:planet_vel} for a range of $K_{\rm p}$ values, and integrated as a function of time/phase. We then plot the CCF at the optimum $K_{\rm p}$ value, an example of which is shown in Fig. \ref{Fe_V_comp}. As the model is unbroadened (i.e., it is not convolved with the broadening kernel, but still includes its natural line widths from thermal broadening) the distribution of the CCF may be used as a proxy for the average line shape in our data, after rotational and instrumental broadening. In other words, the cross-correlation is a convolution of the intrinsic line shape in the data with the unbroadened model. By comparing the shape of the cross-correlation function of Fe, which was previously detected in both limbs \citep{Gandhi_2022,Gandhi_2023}, with that of other species, we can empirically search for any apparent asymmetry in their CCF as a function of $\Delta v_{\text{sys}}$. This asymmetric CCF could potentially be caused by an asymmetric distribution in their abundance between the blueshifted evening limb and redshifted morning limb \citep{Wardenier_2023}. Cross-correlation trails which deviate from the planet's Keplerian velocity curve, such as that first seen in \cite{Ehrenreich_2020}, can also often result in a ``diamond'' feature in the $K_{\rm p}-\Delta v_{\text{sys}}$ maps, as a warped CC-trail summed in time will result in multiple peaks in the map localised at the expected $K_{\rm p}$ and $\Delta v_{\text{sys}}$ values \citep{Nugroho_2020b,Wardenier_2021,Maguire_2023}. Fig. \ref{Fe_V_comp} shows the results of this test for Fe and V, in which we see a narrower CCF for V relative to Fe, for both T1 and T3, caused by a predominantly blueshifted V line shape. 

\section{Discussion}
\label{sec:4}
\subsection{Abundance constraints}
\label{Abuns_discuss}
Our abundance constraints are directly compared with stellar values \citep{Tabernero_2021} where measured  (i.e., Fe, Cr, Mg, Na), as well with solar values \citep{Asplund_2009} where measured. The individual constraints obtained for the morning and evening limbs are also compared with a similar spatially-resolved analysis of WASP-76b, conducted by \citetalias{Gandhi_2023}, for specific species (i.e., Fe, V, Mg, Cr). For species such as Na and VO, where previous spatially-resolved constraints are not available, we directly compare our results with those obtained by \citetalias{Pelletier_2023}. It's important to note that, unlike our approach, \citetalias{Pelletier_2023} modelled the terminator as a whole. The relative abundance constraints for the morning and evening limbs are shown in Figs. \ref{Morn_Relabuns} and \ref{Eve_Relabuns}, respectively. Whereas, the absolute abundance constraints for the morning and evening limbs are shown in Figs. \ref{Morn_abuns} and \ref{Eve_abuns}, respectively.
\subsubsection*{Fe}
Across all three transits, we find consistent Fe abundance constraints in both the morning and evening limbs and find the abundances to be consistent across the limbs of the atmosphere for each of the individual transits. This suggests a stable uniform longitudinal distribution of Fe in the atmosphere of WASP-76b, which is consistent with previous spatially-resolved constraints by \citet{Gandhi_2022,Gandhi_2023}. This uniform distribution of Fe indicates that a global process, such as an uneven cloud deck or temperature asymmetry, is likely responsible for the asymmetric CC trail previously found \citep{Ehrenreich_2020,Savel_2022}. This is further supported by the asymmetric CC trails found for a multitude of species with varying condensation temperatures \citep{Kesseli_2022,Pelletier_2023}.\\ \\ 
\noindent Furthermore, both our morning and evening Fe abundance constraints are consistent with stellar values, which suggests a uniform atmospheric composition similar to that of its parent protoplanetary disk. This is consistent with the findings of \citetalias{Gandhi_2023}, in which Fe/H values of a range of UHJs -- corrected for the stellar metallicity of each system -- were found to be consistent with their host star values. \citetalias{Pelletier_2023} also found a similar trend for the abundances of refractory metals in the atmosphere of WASP-76b.
\subsubsection*{V \& VO}
V is consistent with solar values and \citetalias{Gandhi_2023} for T2. Similarly, VO is also consistent with constraints obtained by \citetalias{Pelletier_2023} for T2 in the morning limb. Both V and VO are unconstrained in the morning limb for both T1 and T3, with upper limits consistent with T2.
\\ \\
\noindent In the evening limb, all three transits are consistent, aligning with solar values as well as with both \citetalias{Gandhi_2023} and \citetalias{Pelletier_2023}. Where constrained, we find V and VO abundances to be consistent across the limbs of the planet.
\\ \\
\noindent There is an interesting discrepancy between the morning and evening constraints for both V and VO. For T2, we constrain both species across both limbs. However, for T1 and T3, we only constrain the species in the evening limb. This leads to the V/Fe ratio in the morning limb for both T1 and T3 (see Fig. \ref{Morn_Relabuns}) to be subsolar. \citet{Gandhi_2022} found that, overall, the combined signal is dominated by the more blueshifted evening limb, which is in agreement with previous GCM predictions \citep{Wardenier_2021}. However, signals from irradiated terminator regions (regions A \& D of \citetalias{Gandhi_2023}) typically dominate the combined signal in their respective phase ranges due to larger temperatures and, thus, scale heights. \citetalias{Gandhi_2023} was also unable to constrain V in the dayside morning region (D). These dominant signals result in tighter constraints for these regions, making it surprising that V is no longer constrained. All three of our observations are of similar quality, with comparable S/N ratios when adjusted for the varying exposure times. We monitored the integrated water vapour (IWV) between observations and found no significant trend which may impact the reliability of our data. This is further supported by the similar constraints shown between each of our injection tests (see Figs. \ref{T1_inj_retrieval} --\ref{T3_inj_retrieval}). We also monitored the variation in airmass between observations and found an increase in airmass for T1 and T3 during ingress, which could explain the dampened morning V/VO signal. However, such an effect would likely reduce the quality of our constraints in the morning limb uniformly across all species, which is not the case, and thus a decrease in data quality for T1 and T3 relative to T2, which might explain worse V/VO constraints in the morning limb, seems unlikely. It is possible that both the V and VO signals are dampened in the cooler morning limb for T1 and T3, due to a combination of recombination and condensation at lower temperatures. However, as our upper limits are consistent with those of T2, as well as \citetalias{Gandhi_2023}, we cannot rule out their presence in the morning limb entirely.\\ \\ \noindent The search for the dampened V signal in the morning limb for both T1 and T3 relative to T2 can be approached more empirically. In Fig. \ref{Fe_V_comp}, we show a slice through the optimal $K_{\rm p}$ value for both Fe and V. The figure shows CCFs of \textit{unbroadened} models that peak at similar $\Delta v_{\rm sys}$ values. Fe has been previously detected in both limbs of the planet by \citet{Gandhi_2022}, as well as in this work. Thus, in order to identify any asymmetries in the line profile of V, we compare the CCFs of Fe and V. For T2, the CCFs of both Fe and V exhibit a similar shape, characterised by a relatively broad profile. This broad shape is due to contributions from both the blueshifted (evening) and redshifted (morning) components of the line profile. In contrast, for both T1 and T3, the CCF of V is notably narrower relative to Fe and is primarily dominated by the blueshifted (evening) component, perhaps indicating a more localised contribution to the line profile.\\ \\
\noindent Although the origin of this asymmetry in the V and VO distributions across the morning and evening limbs for T1 and T3 is unclear, if present, its discrepancy with the well-constrained V and VO abundances measured for both limbs for T2 indicates an atmosphere with clear variable chemistry, in which a combination of recombination/condensation is altering the V and VO abundances at the cooler morning limb over the course of months/years.
\\ \\ 
\noindent Although the VO line list used (VOMYT; \cite{McKemmish_2016}) has resulted in detections of VO in the atmosphere of WASP-76b (e.g., \citetalias{Pelletier_2023} \& Fig. \ref{VO_comp}), it has previously led to non-detections of VO in the atmosphere of WASP-121b at high-resolution \citep{Hoeijmakers_2020, Merritt_2020}. These non-detections were reported as inconclusive, and this line list was shown to be insufficient in recovering injected VO signals from UVES transmission spectroscopy data of WASP-121b \citep{de_Regt_2022}. Recently, \cite{Bowesman_2024} presented an updated MARVEL-ised \citep[Measured Active Rotational-Vibrational Energy Levels;][]{CSASZAR_2007} line list for VO (HyVO), which includes hyperfine splitting effects and should facilitate more reliable high-resolution detections and abundance constraints of VO. 
\\ \\ 
We have conducted a cross-correlation analysis with the updated line list, with cross-sections computed at $T=2800$ K and $P=0.1$ mbar, shown in Fig. \ref{HyVO_VOMYT}. In order to determine the optimum parameters for this cross-correlation analysis, we ran an MCMC similar to that outlined in Sect. \ref{sec:Retrieval}, using a simple one-dimensional Gaussian broadening kernel, and an isothermal $T$-$P$ profile. We have now detected VO in T1 via cross-correlation. However, our previous detection of VO in T2 has now shifted to a significantly lower $\Delta v_{\rm sys}$ and $K_{\rm p}$. Our T3 detection significance has increased with the update line list. It is difficult to assess the accuracy of these new detections/non-detections without performing a full atmospheric retrieval, with a varying $T$-$P$ profile and two-dimensional broadening kernel. As the updated cross-sections are not readily available across our full $T$-$P$ grid, an atmospheric retrieval akin to that presented above is impracticable at this time. A quantitative assessment of the updated line list is beyond the scope of this work; however, it is required to ensure its reliability for exoplanetary studies at high-resolution going forward. Despite this added uncertainty in terms of a VO detection, our results still show a potentially variable V abundance, with which there is no uncertainty regarding line list accuracy and, alongside this VO analysis, provides an interesting insight into the chemistry of V-bearing species.
\subsubsection*{Cr}
The Cr/Fe constraints obtained are also consistent between transits, as well as with \citetalias{Gandhi_2023}, for both the morning and evening limbs. Similarly to V and VO, however, we find Cr/Fe to be depleted relative to stellar for T1 and T3 in the morning limb, potentially due to partial recombination/condensation of Cr to other compounds, such as CrH, which has previously been detected in the atmosphere of WASP-31b \citep{Braam_2021,Flagg_2023}. \citetalias{Pelletier_2023} also constrain the abundance of CrH in their 1D retrieval of WASP-76b, however were unable to detect it via cross-correlation. Both \citetalias{Pelletier_2023} and \citetalias{Gandhi_2023} also found substellar Cr/Fe ratios, within 1$\sigma$, consistent with our findings.\\ \\ \noindent We compare relative abundances as these constraints are typically more accurate and reliable at high-resolution, and in general for transmission spectra. This is due to the well-known degeneracy between the absolute abundance, $\chi_{\rm species}$, and opaque cloud deck pressure, $P_{\rm cloud}$, and/or reference pressure, such that a model with a higher $\chi_{\rm species}$ and lower $P_{\rm cloud}$ could, in principle, produce identical features as a model with a lower $\chi_{\rm species}$ and higher cloud deck pressure $P_{\rm cloud}$ \citep{Benneke_Seager_2012}. Furthermore, the normalisation of high-resolution spectra during preprocessing means we are only sensitive to line strengths above the planetary continuum, making relative line strengths, and consequently, relative abundance constraints, more precise \citep{Birkby_2018,Gibson_2020}. 
% However, Gandi2020c.
\subsubsection*{Mg}
Similarly to Fe, our Mg abundance constraints are consistent with both stellar values and \citetalias{Gandhi_2023} across all three transits. They are also consistent across the limbs of the atmosphere for each of the individual transits. Interestingly, the absolute abundance of Mg is consistent with the upper prior limit (see Tab. \ref{tab:Table2}) for both the morning (T1 and T2) and evening (T2 and T3) limbs. Similar behaviour has been seen previously in high-resolution analysis of WASP-121b with UVES and ESPRESSO \citep{Gibson_2022,Maguire_2023}. As Mg II has previously been detected in the upper atmosphere of WASP-121b \citep{Sing_2019}, it is thought that Mg may be part of an extended, non-hydrostatic envelope, such that our model assumption of hydrostatic equilibrium breaks down. To account for this, our model would artificially increase the abundance of Mg, increasing its line strengths. It is difficult to say if this is also the case for WASP-76b, as Mg II has not yet been detected in its atmosphere. Furthermore, Mg has relatively few lines in the ESPRESSO wavelength range, which could also explain its poorer abundance constraints.
\subsubsection*{Na}
We find a substellar Na abundance relative to Fe in the evening limb across all three transits. This is in agreement with \citetalias{Pelletier_2023}, who find substellar abundances for alkali metals (Li, Na, K), suggesting their lower ionisation energies cause a greater degree of ionisation at the temperatures and pressures probed, driving a decrease in the abundance of neutral alkaline atoms. Similar behaviour was found in the atmosphere of WASP-121b, where the abundance of both Na and Ca relative to Fe were typically lower than their stellar counterparts, alongside the presence of high-altitude Ca II indicates a greater degree of ionisation \citep{Maguire_2023}. It is also worth noting that \citetalias{Gandhi_2023} were unable to constrain Na in both the morning and evening limbs of WASP-76b via an independent joint retrieval of T1 and T2. \\ \\ \noindent The morning Na abundances relative to Fe were also found to be substellar for both T1 and T2, consistent with the evening limb and continuing the trend mentioned above. However, for T3, we find a drastically higher Na/Fe ratio driven by an anomalously high Na abundance. It is possible that Na is also part of an extended outer envelope, similar to Mg, becoming more readily ionised in the upper atmosphere, which is accounted for on the morning limb by an artificial increase in abundance. Another possible explanation is the existence of a high-altitude cloud deck on the morning limb, or similar continuum opacity source such as H$^-$ opacity, which is not accounted for in our model. This drives the Na abundance, creating a pseudo-continuum. In order to ensure this artificially high Na abundance does not affect the abundances of the other species in our T3 retrieval, the retrieval was re-run without the presence of Na. We find it has a minimal impact on our retrieved abundances, however, alters our morning $T$-$P$ profile constraints significantly. This is to be expected, as a higher Na abundance with a smaller uncertainty will result in a lower temperature with tighter constraints.
\subsection{Temperature constraints}
For T1, our morning and evening temperatures at $0.1$ mbar, $T_{0.1\rm mbar}$, vary at the 1$\sigma$ level (see Tab. \ref{tab:Table3} and Fig. \ref{T1_retrieval}). These morning and evening temperature constraints are consistent with the spatially-resolved temperature constraints of \citet{Gandhi_2022}. Above the high-altitude morning cloud deck, we find a morning $T$-$P$ profile which is cooler than that of the evening limb, which is inverted above $\sim$0.1 mbar.\\ \\ \noindent For T2, we find a cooler evening $T$-$P$ profile which varies from that of the morning limb at the 2$\sigma$ level at 0.1 mbar (see Tab. \ref{tab:Table3} and Fig. \ref{T2_retrieval}). Our evening temperature at $0.1$ mbar is consistent with that of \citet{Gandhi_2022}. For T2, the $T_{0.1\rm mbar}^{\rm Morn}$ is significantly higher than that of \citet{Gandhi_2022}, and at high pressures, it reaches unphysically high values for WASP-76b. 
% This adverse increase in temperature could be due to our model neglecting the presence of H$^-$ opacity, whose presence is known to cool the atmosphere in this pressure regime \citep{Parmentier_2018}. 
This discrepancy between the morning limb with both previous transits presented in this study, as well as with previous studies, is also likely explained by the presence of V and VO in the morning limb, whereas it's unconstrained for T1 and T3, as VO is known to be a thermal inversion agent which drives high-altitude temperature inversions in the upper atmospheres of UHJs. It is worth noting that transmission spectroscopy studies are typically not very sensitive to the $T$-$P$ structure of the atmosphere, and therefore we must be cautious not to overinterpret our constraints.
\begin{table}[!htbp]
    \centering
    \caption{Temperature constraints for T1, T2, and T3 compared to \citetalias{Gandhi_2023}.}
    \label{tab:Table3}
    \begin{tabularx}{\columnwidth}{*{3}{>{\centering\arraybackslash}X}} % Two columns, centered
        \hline
        \hline \\[-0.3cm]
        Transit & $T_{0.1\rm mbar}^{\rm Morning}$ (K) & $T_{0.1\rm mbar}^{\rm Evening}$ (K)\\ [0.1cm]
        \hline \\[-0.3cm]        % Add your table content here
        T1 & $2820^{+140}_{-50}$ & $2240^{+530}_{-170}$ \\[0.15cm]
        T2 & $3650^{+120}_{-60}$ & $3060^{+180}_{-70}$ \\[0.15cm]
        T3 & $1360^{+130}_{-60}$ & $2120^{+810}_{-460}$ \\[0.15cm]
        \citetalias{Gandhi_2023} & $2615^{+266}_{-275}$ & $2950^{+111}_{-156}$  \\[0.15cm]
        \hline
           
    \end{tabularx}
\end{table}
\\ \\ \noindent For T3 (see Tab. \ref{tab:Table3} and Fig. \ref{T3_retrieval}), we find a hotter evening $T$-$P$ profile which is inverted above $\sim$0.1 mbar, similar to T1. The evening temperature at $0.1$ mbar is consistent with that of \citet{Gandhi_2022}, and varies from the morning limb within 1$\sigma$. The $T_{0.1\rm mbar}^{\rm Morn}$ is significantly lower than that of \citet{Gandhi_2022}. This adverse decrease in temperature could be due to our model neglecting the presence of H$^-$ opacity, whose presence is known to cool the atmosphere in this pressure regime \citep{Parmentier_2018}. As mentioned in Sect. \ref{Abuns_discuss}, this temperature discrepancy is also driven by the presence of an anomalously high Na abundance on the morning limb of T3. When we exclude Na from our retrievals, we find a much broader morning temperature constraint of $T_{0.1\rm mbar}^{\rm Morn} = 1950^{+990}_{-380}$ K, which is consistent with that found in \citet{Gandhi_2022}. The evening $T$-$P$ profile does not change significantly.
\subsection{Atmospheric dynamics}
 For T1, we retrieve $K_{\rm p} = 185.48\pm1.22$ km/s and $\Delta v_{\rm sys} = -7.70\pm0.82$ km/s, for T2: $K_{\rm p} = 175.48\pm1.33$ km/s and $\Delta v_{\rm sys} = -8.77\pm0.44$ km/s, and for T3: $K_{\rm p} = 181.33\pm2.14$ km/s and $\Delta v_{\rm sys} = -7.88\pm0.66$ km/s. Although we constrain different values for each of our transits, we compute the planetary orbital velocity, given Eq. \ref{eq:planet_vel}, using 10,000 random samples from our posterior distributions of $K_{\rm p}$ and $\Delta v_{\rm sys}$, and find $v_{\rm p}$ consistent within 1$\sigma$ in-transit. There is also a degeneracy between the global systemic velocity offset, $\Delta v_{\rm sys}$, and the blueshifted component of the rotational velocity, $\Delta v_{\text{rot}, e}$, with the latter ideally accounting for subtle variations in the individual line shapes caused by additional wind patterns and rotation. When accounting for a broadening of an atmospheric model with R=200,000, we find broadening parameters of $v_{\rm rot, m} = 1.24\pm1.16$ km/s and $v_{\rm rot, e} = 2.97\pm1.61$ km/s for T1. We also determine $v_{\rm rot, m}$ and $v_{\rm rot, e}$ for T2 as $2.50\pm1.45$ km/s and $0.99\pm0.72$ km/s, respectively. For T3, we find $v_{\rm rot, m}$ and $v_{\rm rot, e}$ as $5.77\pm2.62$ km/s and $1.33\pm0.94$ km/s. It's important to note that these distributions are non-Gaussian, and therefore, the constraints may not be accurately represented by the median value and standard deviation alone. The combination of these broadening parameters with a constant blueshifted offset of $\sim$8 km/s is in agreement with previous high-resolution studies of WASP-76b \citep{Ehrenreich_2020,Pelletier_2023,Gandhi_2022,Gandhi_2023}. \\ \\ \noindent The treatment described above also assumes a single $K_{\rm p}$, $\Delta v_{\rm sys}$, $\Delta v_{\text{rot}, m}$, and $\Delta v_{\text{rot}, e}$ value governs the dynamics of each of our atmospheric species. In reality, as seen in previous high-resolution studies \citep{Prinoth_2022,Wardenier_2023}, velocity offsets \textit{between} species can occur. In reality, each species is likely to exist in different dynamical regimes, particularly when considering differing atmospheric limbs and/or an extended upper atmosphere. A more robust modelling approach would involve the individual treatment of each species' dynamics (i.e, varying $K_{\rm p}$, $\Delta v_{\rm sys}$, $\Delta v_{\text{rot}, m}$, and $\Delta v_{\text{rot}, e}$ for each species), which we leave as an avenue for future research.\\ \\ 
Recent studies using low-resolution transmission spectroscopy with JWST have also allowed us to investigate the three-dimensional nature and varying chemistry across the limbs of these objects. For example, \cite{Rustamkulov_2023} found a wavelength-dependent central transit time for WASP-39b, suggesting an asymmetric distribution of atmospheric species across the terminator. Furthermore, it has been shown that by probing the varying transit lightcurves of each atmospheric limb, particularly during ingress and egress, one can retrieve separate transmission spectra for each half of the terminator \citep{Espinoza_2021,Grant_2022}. These techniques at low-resolution, coupled with analysis such as that provided in this study, highlight the enormous promise of utilising low- and high-resolution transmission spectroscopy observations to investigate the three-dimensional nature of close-in exoplanets, a regime which until now was accessible only with phase curve measurements.
\section{Conclusions}
\label{sec:5}
We have presented multiple high-resolution transit observations of the atmosphere of the UHJ WASP-76b, monitored across several months/years, with the ESPRESSO spectrograph at the VLT. In this study, we have modelled the morning (leading) and evening (trailing) limbs of the atmosphere separately with the use of realistic rotational broadening kernels, allowing us to retrieve separate atmospheric abundances, $T$-$P$ profiles, and dynamical parameters for each atmospheric limb via our atmospheric retrieval framework.
\begin{itemize}
    \item We have confirmed the presence of VO, recently discovered in the atmosphere of WASP-76b by \cite{Pelletier_2023}, via cross-correlation in both T2 and T3 of our observations, with a detection significance of 4.0$\sigma$ and 7.2$\sigma$, respectively.\\
    \item We have found a uniform longitudinal distribution of Fe and Mg across the limbs of the atmosphere, for each of our transits, consistent with previous works as well as stellar abundances.\\
    \item We constrained Cr/Fe and Na/Fe ratios which are broadly consistent between the limbs of the atmosphere for each of the transits, with the exception of Na/Fe in the morning limb for T3. These abundance ratios were also found to be substellar in both the morning and evening limbs, hinting at possible depletion of Cr and Na due to recombination and ionisation, respectively, as suggested in previous works.\\
    \item The V/Fe and VO/Fe ratios were also found to be consistent between the limbs of the atmosphere for each of the transits, with V/Fe also consistent with solar in the evening limb. V and VO were, however, unconstrained in the morning limb for T1 and T3, suggesting possible depletion due to recombination and condensation. This is further supported by an apparent asymmetric CCF for V relative to Fe for both T1 and T3.
\end{itemize}
The consistent results obtained across multiple observations spanning months or years, as well as with previous studies using different modelling and retrieval frameworks and/or instruments, affirms the efficacy of high-resolution atmospheric retrieval frameworks. Spatially-resolved temporal atmospheric surveys such as this, coupled with low-resolution efforts in the era of JWST, will allow us to place
further constraints on the time-varying dynamics and chemistry of these worlds, allowing us to infer how they evolved to their current climate.
\begin{acknowledgements}
This work relied on observations collected at the European Organisation for Astronomical Research in the Southern Hemisphere. The author(s) gratefully acknowledge support from Science Foundation Ireland and the Royal Society in the form of a Research Fellows Enhancement Award No. RGF\textbackslash EA\textbackslash 180315. SKN is supported by JSPS KAKENHI grant No. 22K14092. We are grateful to the developers of the NumPy, SciPy, Matplotlib, iPython, corner, and Astropy packages, which were used extensively in this work \citep{Numpy,Scipy,matplotlib,ipython,corner,astropy}.
\end{acknowledgements}

% % WARNING
% %-------------------------------------------------------------------
% % Please note that we have included the references to the file aa.dem in
% % order to compile it, but we ask you to:
% %
% % - use BibTeX with the regular commands:
\bibliographystyle{aa} % style aa.bst
\bibliography{sample.bib} % your references Yourfile.bib

% Appendix A

\begin{appendix} %First appendix

\section{Derivation of the broadening kernel}
\label{appendix_a}
\begin{figure}
\resizebox{\hsize}{!}{\includegraphics{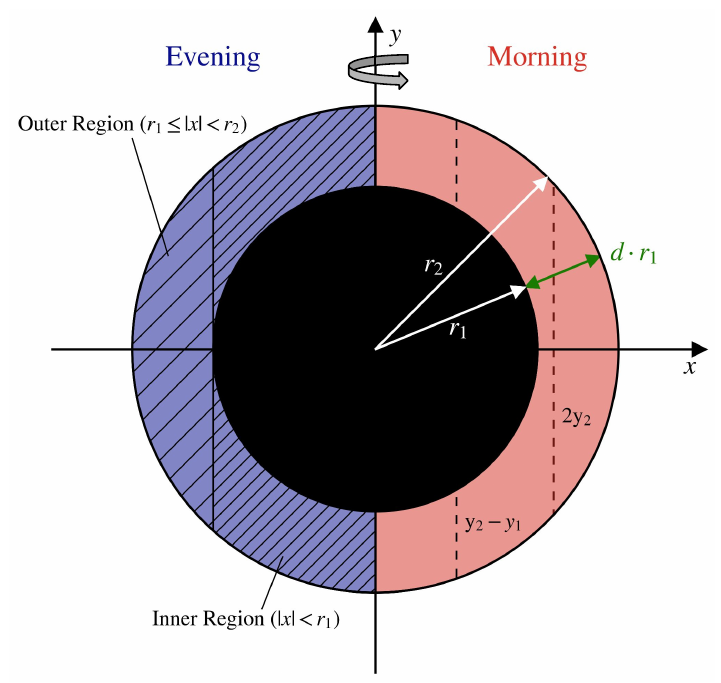}}
\caption{Schematic diagram of a rotating annulus of inner radius $r_1$ and outer radius $r_2 = r_1 + d\cdot r_1$.  A portion of the inner and outer regions described in the text is highlighted. Chords at constant x, of length $2y_2$ and $2\hspace{0.3mm}(y_2 - y_1)$, are also shown.}
\label{Annulus}
\end{figure}

For a uniformly irradiated disk projected by a rigid rotating sphere, the broadening profile $G(\Delta v)$ is given by:
\begin{align}
\label{eqn:A1}
    G(\Delta v) &= \frac{I_\lambda(\Delta v)}{I_\lambda(0)}
\end{align}
where $I_\lambda(0)$ is the line intensity at the line centre and $I_\lambda(\Delta v)$ is the line intensity after a Doppler shift caused by the line of sight velocity, $\Delta v$, at each $x$ coordinate along the disc. For a rigid rotating sphere, such that there is no latitudinal change in rotation rate, all elements on the disk surface along the same $x$ coordinate undergo the same Doppler shift, such that $x = \Delta v / \Delta v_{\text{rot}}$, where $\Delta v_{\text{rot}}$ is the Doppler shift due to the equatorial velocity ($r = r_1$). Eq. \ref{eqn:A1} is given by the ratio of the length of the chord at constant $x$, to the length of the chord at $x = 0$.\\ \\
\noindent This expression can be adapted for a uniformly irradiated annulus projected by a rigid rotating spherical shell (see Fig. \ref{Annulus}), such as a planetary atmosphere. Fig \ref{Annulus} shows a rotating annulus with inner radius $r_1 = R_{\rm p}$. We set $r_1 = R_{\rm p} = 1$, such that the outer radius $r_2 = 1 + d$, where $d$ denotes the thickness of the atmosphere \textit{relative} to the planetary radius $R_{\rm p}$. Thus, the length of the chord at $x = 0$ is $2d$. See \cite{Boucher_2023} for a similar expression with alternative notation, in which the absolute height of the atmosphere $z=d\cdot R_{\rm p}$. The length of the chord at constant $x$ varies for different regions of the annulus.
\subsection*{\textbf{Outer region ( \boldmath{$r_1 \leq \lvert x \rvert < r_2$})}}
When $r_1 \leq \lvert x \rvert < r_2$, the outer disk ($r = r_2$) is not occulted by the planetary disc, such that the lines simply undergo rotational broadening due to a rigid rotating disk ranging from $-r_2 \rightarrow r_2$. Therefore the length of the chord at constant $x$ is simply:
\begin{align*}
    2y_2 &= 2\sqrt{r_2^2 - x_2^2} \\
    &=  2\sqrt{(1+d)^2 - x_2^2}
\end{align*}
As all elements on the disk surface along the same x
coordinate undergo the same Doppler shift, this expression becomes:
\begin{align*}
    2y_2 &=  2\sqrt{(1+d)^2 - (\Delta v/\Delta v_{\text{rot}})^2}\\
\end{align*}
Such that Eq. \ref{eqn:A1} is:
\begin{align*}
    G(\Delta v) &= \frac{2y_2}{2d}\\
    &= \frac{\sqrt{(1+d)^2 - (\Delta v/\Delta v_{\text{rot}})^2}}{d} \numberthis
\end{align*}
\subsection*{\textbf{Inner region ( \boldmath{$\lvert x \rvert < r_1$})}}
However, when $\lvert x \rvert < r_1$, the outer disk ($r = r_2$) is occulted by the inner disk ($r = r_1$), such that the lines undergo rotational broadening due to a rotating spherical shell of inner radius $r_1$ and outer radius $r_2$. The length of the chord at constant $x$ is:
\begin{align*}
    2\hspace{0.3mm}(y_2 - y_1) &= 2\hspace{0.3mm}\left(\hspace{-0.8mm}\sqrt{(1+d)^2 - (\Delta v/\Delta v_{\text{rot}})^2} - \sqrt{1 - (\Delta v/\Delta v_{\text{rot}})^2}\hspace{0.8mm}\right)
\end{align*}
Such that 
\begin{align*}
    G(\Delta v) &= \frac{2\hspace{0.3mm}(y_2 - y_1)}{2d}\\
    &= \frac{\sqrt{(1+d)^2 - (\Delta v/\Delta v_{\text{rot}})^2} - \sqrt{1 - (\Delta v/\Delta v_{\text{rot}})^2}}{d} \numberthis
\end{align*}
% From \cite{Gray_1992}, Eq. 18.11, for a disk projected by a rigid rotating sphere, such that the Doppler shift is constant for all $y$, the broadening profile of the stellar lines with specific intensity $I_c$ is given by:
% \begin{align}
% \label{eqn:A1}
% G(\Delta v)=
% \begin{dcases}
%     \frac{1}{\Delta v_{\text{rot}}}\cdot\frac{\int_{-y}^{y}I_c \frac{dy}{R}}{\oint I_c \cos{\theta} d\omega}& \text{, for } \lvert \Delta v \rvert \leq \Delta v_{\text{rot}}\\
%     \hfil 0 & \text{, for } \lvert \Delta v \rvert > \Delta v_{\text{rot}}
% \end{dcases} 
% \end{align}
% where $\Delta v$ and $\Delta v_{\text{rot}}$ are the line of sight velocity at each $x$ coordinate and the equatorial velocity, respectively. $R$ is the stellar radius, $\theta$ is the angular limb distance, $\epsilon$ is the limb darkening coefficient, $I_c^0$ is the specific intensity at the center of the disc, and $d\omega = \frac{dA}{R^2}$ where $dA$ is an incremental surface area on the projected disc. $G(\Delta v) = 0$ for $|\Delta v| > \Delta v_{\text{rot}}$, as we are outside the radius of the sphere. \\

% \noindent The specific intensity, $I_c$, is:
% \begin{align*}
% I_c &= I_c^0 (1-\epsilon\cos{\theta})
% \end{align*}
% which, assuming a uniformly irradiated disk ($\epsilon = 0$), is simply the specific intensity at the center of the disc:
% \
\section{Stellar parameters}

\begin{table}[!htbp]
	\centering
	\caption{Physical parameters of WASP-76. Values adopted from \cite{Ehrenreich_2020}.}
	\label{tab:TableA1}
	\begin{tabular}{cc} % four columns, alignment for each
		\hline
		\hline
				Parameter & Value \\
		\hline
		\hline
    \multicolumn{1}{c}{} & \multicolumn{1}{c}{} \\[-0.2cm]
    $R_\star$ & 1.756$\pm$0.071 R$_{\odot}$\\[0.1cm]
    $M_\star$ & 1.458$\pm$0.021 M$_{\odot}$\\[0.1cm]
    $T_{\rm eff}$ & 6329$\pm$65  K \\[0.1cm]
    Spectral type & F7V \\[0.1cm]
    $\log{g}$ [cgs] & 4.196$\pm$0.106 \\[0.1cm]
    [Fe$/$H] & 0.366$\pm$0.053 \\[0.1cm]
    $K_\star$ & 116.02$^{+1.29}_{-1.35}$ m s$^{-1}$ \\[0.1cm]
    $v_\star\sin{i}$ & 1.48$\pm$0.28 km s$^{-1}$ \\[0.1cm]
    $v_{\rm sys}$ & -1.11$\pm$0.050 km s$^{-1}$ \\[0.1cm]
    $\lambda$ & 61.28$^{+7.61}_{-5.06}$ deg \\[0.1cm]
    \hline
    \end{tabular}

\end{table}
\clearpage
\onecolumn
\section{Injection \& recovery tests}
\label{appendix_b}
\begin{figure*}[!htbp]
    \centering
    \includegraphics[width=\textwidth]{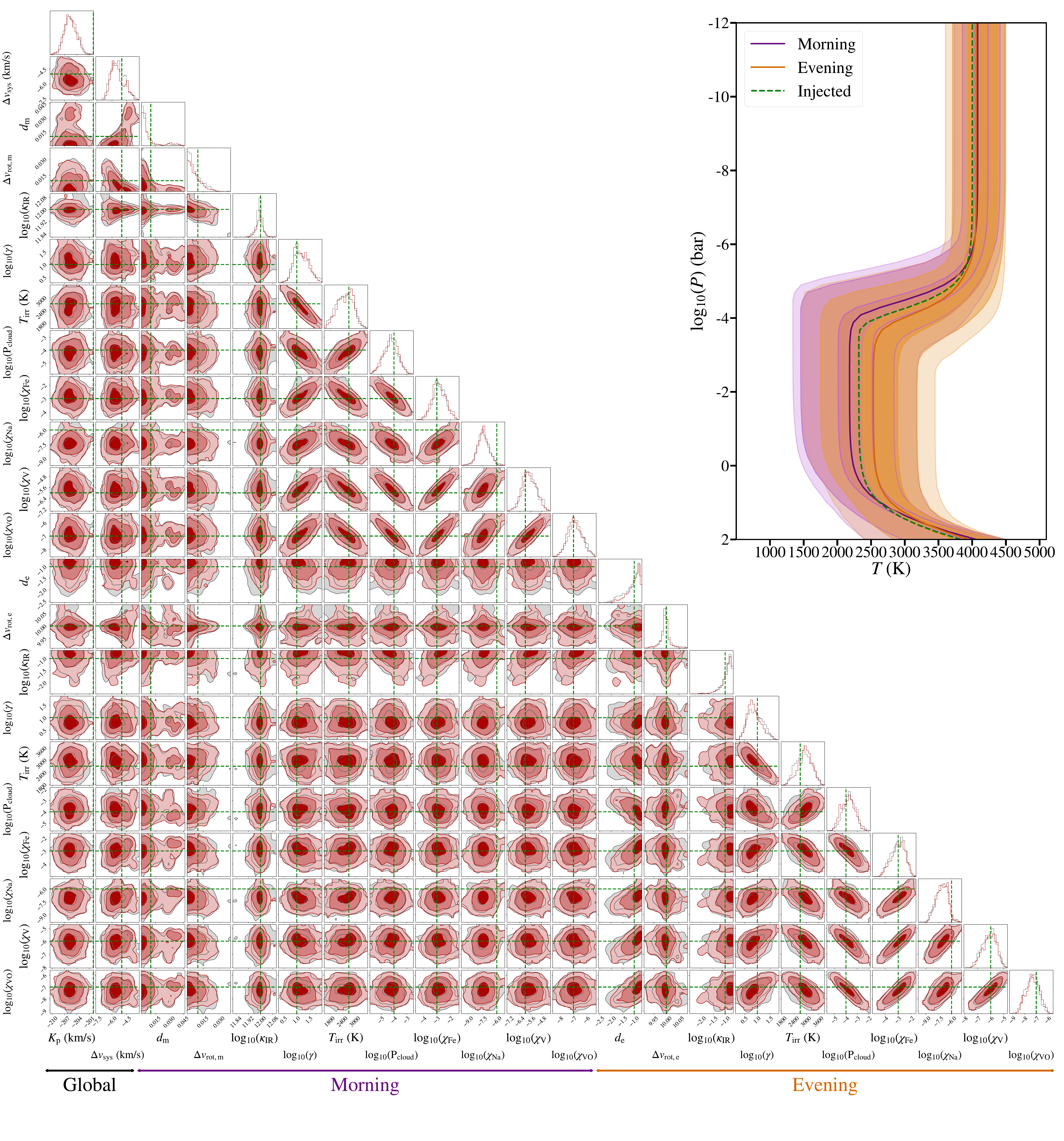}
        \caption{A summary of our injection and recovery tests for T1 outlined in Sect. \ref{sec:Inj_Rec}, with the 1D and 2D marginalised posterior distributions of each of our model parameters displayed. The red and grey posterior distributions represent independent subchains, of the same MCMC chain, both converging to similar distributions, with the injected values highlighted by horizontal/vertical dashed green lines. The various global, morning, and evening parameters are highlighted. The $T$-$P$ profiles on the right were computed from 10,000 random samples of the MCMC, where the solid curve shows the median profile, and the shaded regions show the 1$\sigma$, 2$\sigma$, and 3$\sigma$ contours, for both the morning and evening limbs. The dashed green curve shows the injected $T$-$P$ profile. This corner diagram was generated using \textsc{corner} \citep{corner}.}
        \label{T1_inj_retrieval}
\end{figure*}
\begin{figure*}[!htbp]
    \centering
    \includegraphics[width=\textwidth]{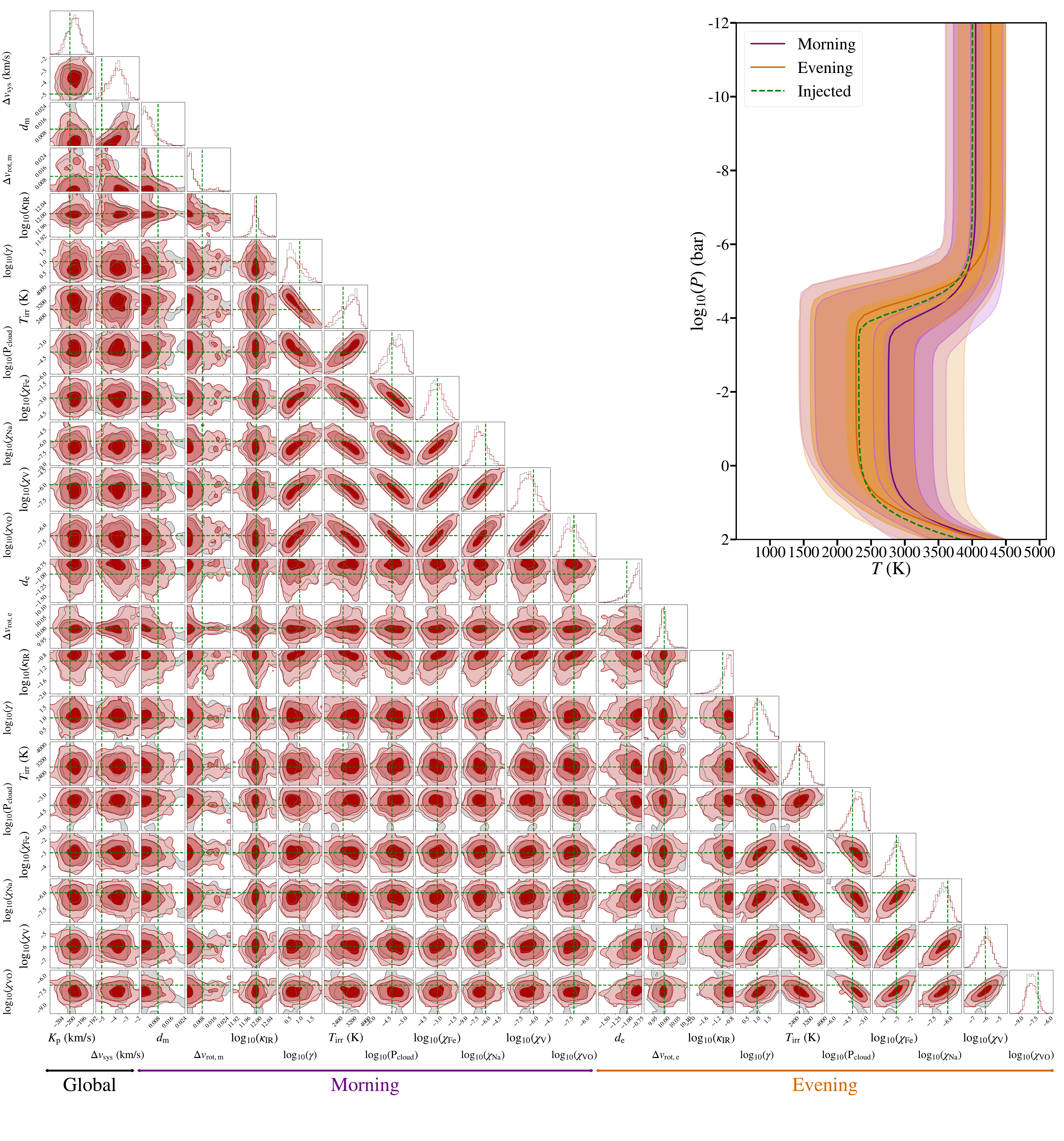}
        \caption{Similar to Fig. \ref{T1_inj_retrieval}, for T2.}
        \label{T2_inj_retrieval}
\end{figure*}
\begin{figure*}[!htbp]
    \centering
    \includegraphics[width=\textwidth]{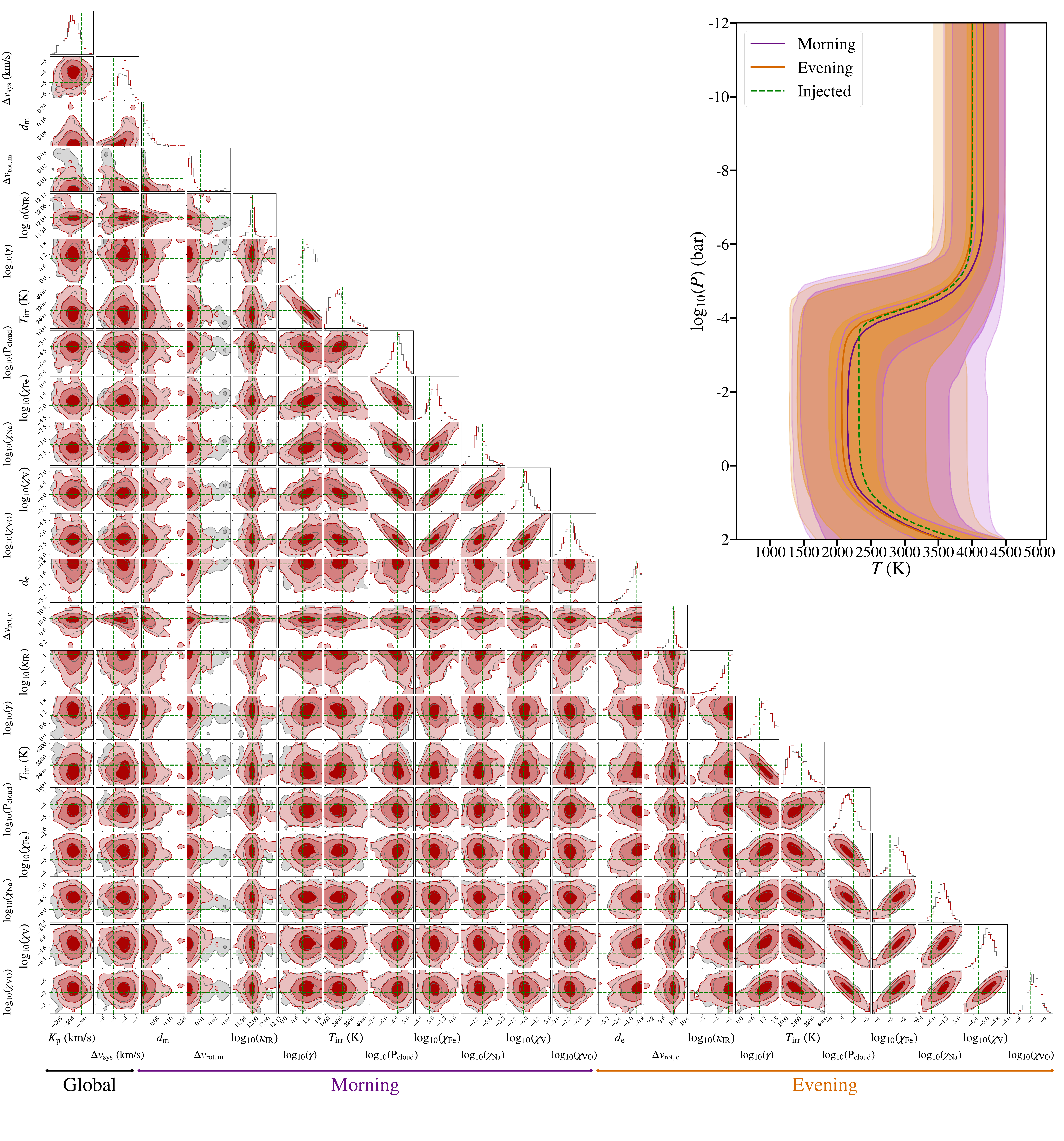}
        \caption{Similar to Fig. \ref{T1_inj_retrieval}, for T3.}
        \label{T3_inj_retrieval}
\end{figure*}
\clearpage
\onecolumn
\section{Atmospheric retrievals}
\label{appendix_c}
\begin{figure*}[!htbp]
    \centering
    \includegraphics[width=\textwidth]{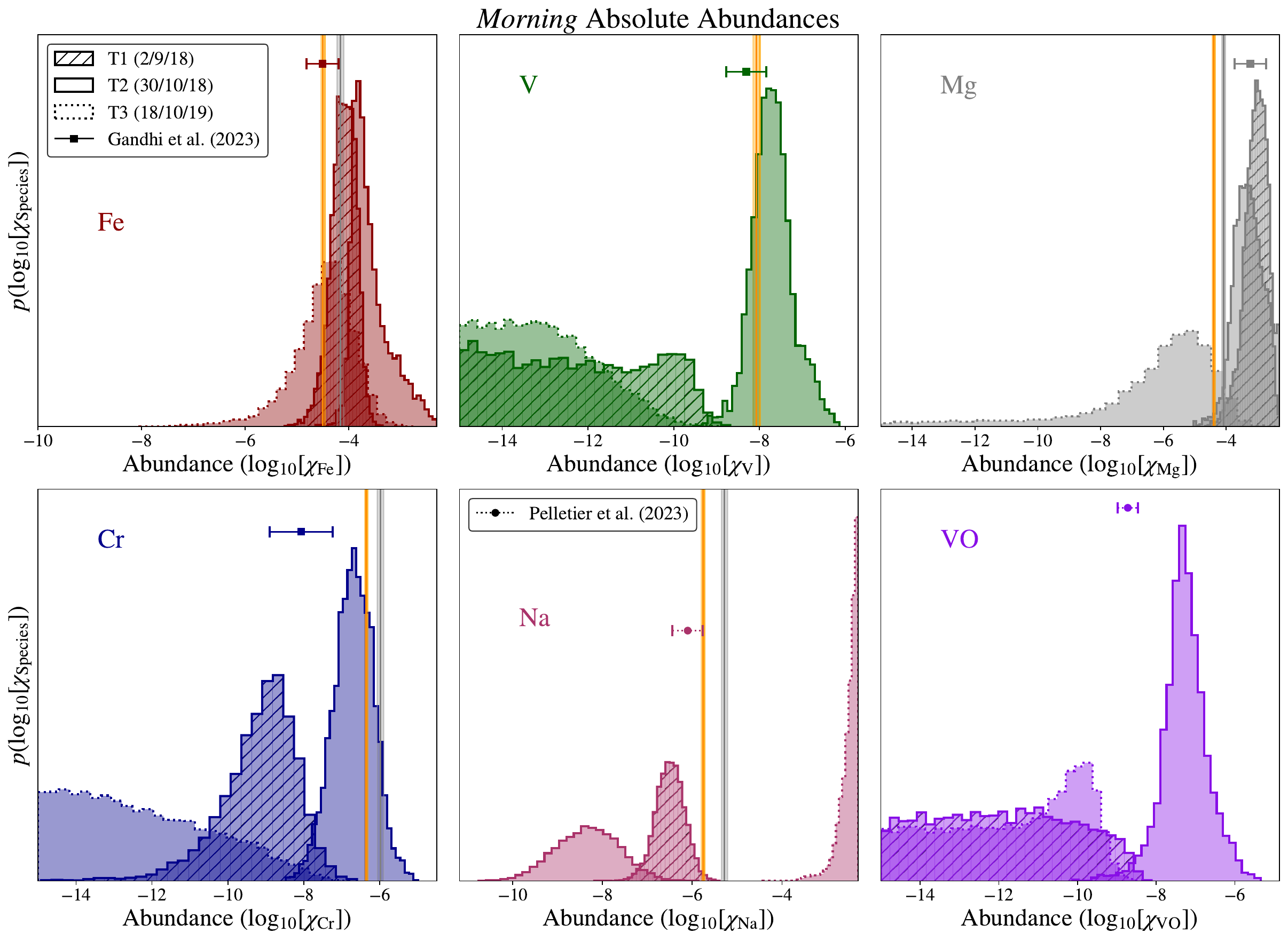}
        \caption{Absolute abundance comparison across each of the three transits, for the \textit{morning} limb. The stellar values from \cite{Tabernero_2021} are shown as a grey vertical line with their $1\sigma$ contours shaded, where measured. The solar values from \cite{Asplund_2009} are shown as an orange vertical line with their $1\sigma$ contours shaded. The weighted mean of the retrieved abundances in regions C and D (morning) from \citetalias{Gandhi_2023} are also shown, where measured. Similarly, the retrieved abundances for the whole terminator from \citetalias{Pelletier_2023} are also shown for Na and VO.}
        \label{Morn_abuns}
\end{figure*}
\begin{figure*}[!htbp]
    \centering
    \includegraphics[width=\textwidth]{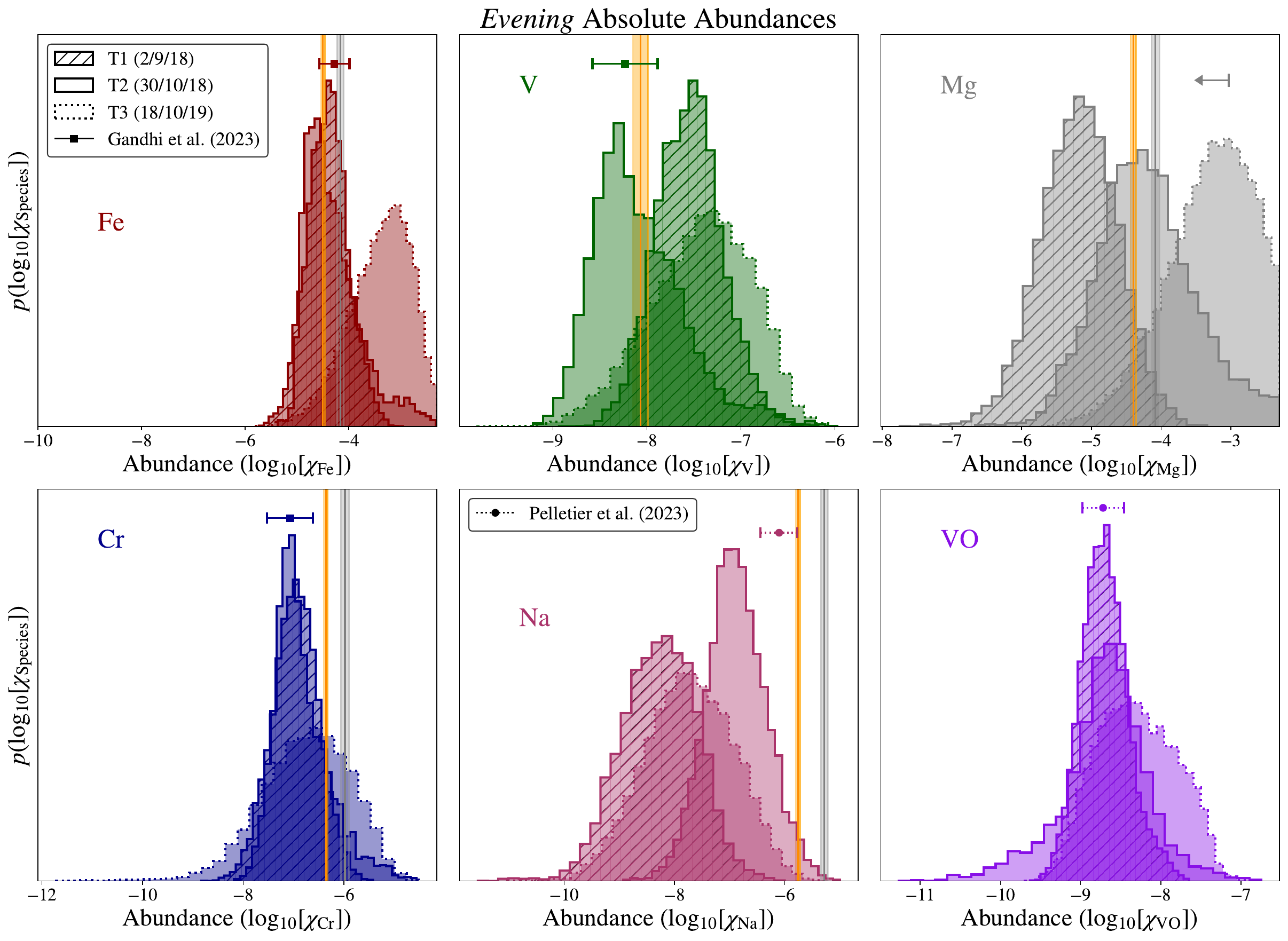}
        \caption{Absolute abundance comparison across each of the three transits, for the \textit{evening} limb. The shaded regions are as in Fig. \ref{Morn_abuns}. The weighted mean of the retrieved abundances in regions A and B (evening) from \citetalias{Gandhi_2023} are also shown, where measured.}
        \label{Eve_abuns}
\end{figure*}
\begin{figure*}[!htbp]
    \centering
    \includegraphics[width=\textwidth]{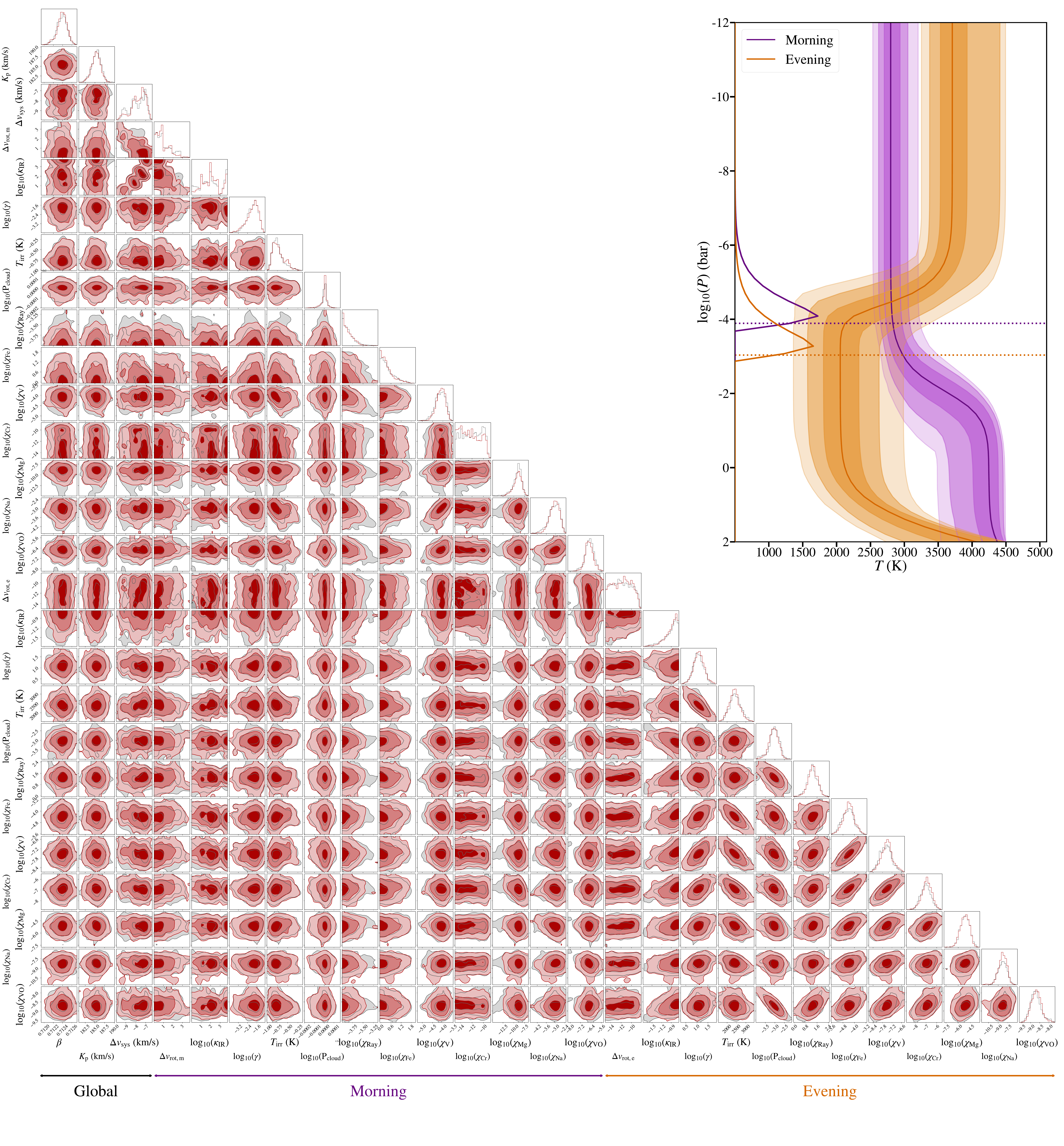}
        \caption{A summary of our atmospheric retrieval results for T1 outlined in Sect. \ref{sec:Retrieval}, with the 1D and 2D marginalised posterior distributions of each of our model parameters displayed. The red and grey posterior distributions represent independent subchains, of the same MCMC chain, both converging to similar distributions. The various global, morning, and evening parameters are highlighted. The $T$-$P$ profiles on the right were computed from 10,000 random samples of the MCMC, where the solid curve shows the median profile, and the shaded regions show the 1$\sigma$, 2$\sigma$, and 3$\sigma$ contours, for both the morning and evening limbs. The horizontal dotted lines are the median retrieved cloud deck pressures, $\log_{10}(P_{\rm cloud})$, below which the model spectrum is truncated. Also plotted are the mean contribution functions for the ESPRESSO wavelength range. This corner diagram was generated using \textsc{corner} \citep{corner}.}
        \label{T1_retrieval}
\end{figure*}
\begin{figure*}[!htbp]
    \centering
    \includegraphics[width=\textwidth]{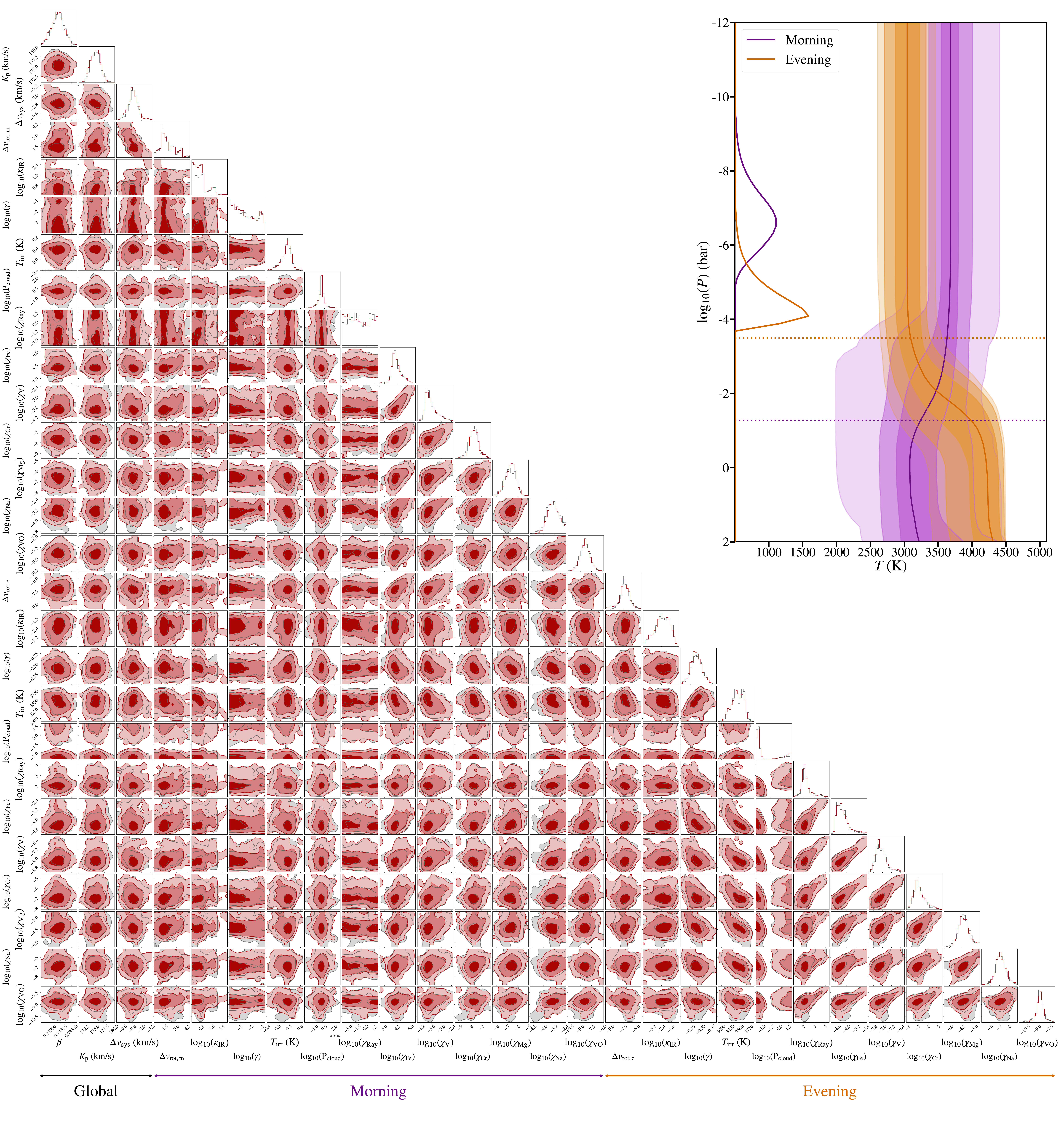}
        \caption{Similar to Fig. \ref{T1_retrieval}, for T2.}
        \label{T2_retrieval}
\end{figure*}
\begin{figure*}[!htbp]
    \centering
    \includegraphics[width=\textwidth]{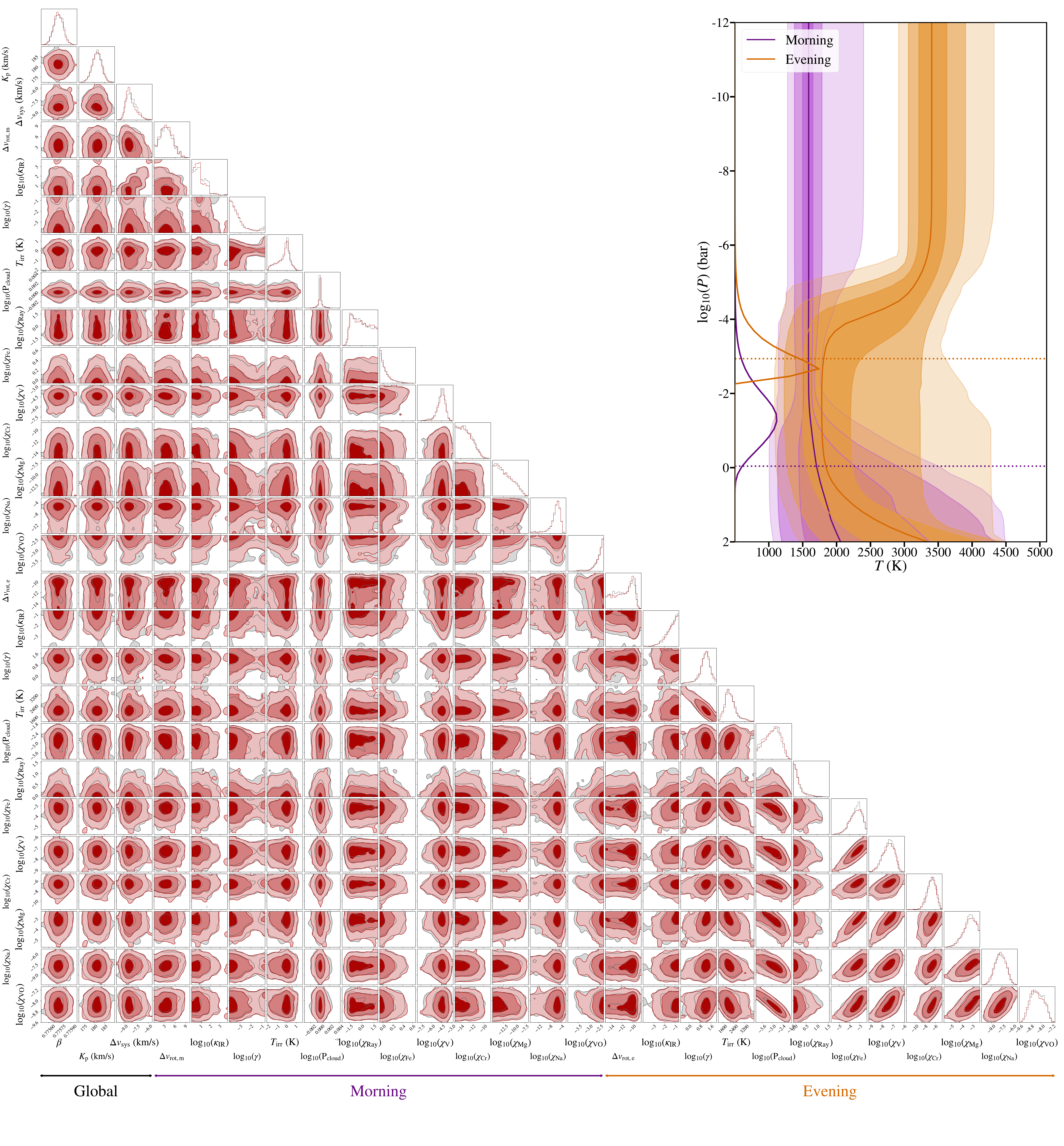}
        \caption{Similar to Fig. \ref{T1_retrieval}, for T3.}
        \label{T3_retrieval}
\end{figure*}
\clearpage
\onecolumn
\section{Cross-correlation}
\begin{figure*}[!htbp]
    \centering
    \includegraphics[width=\textwidth]{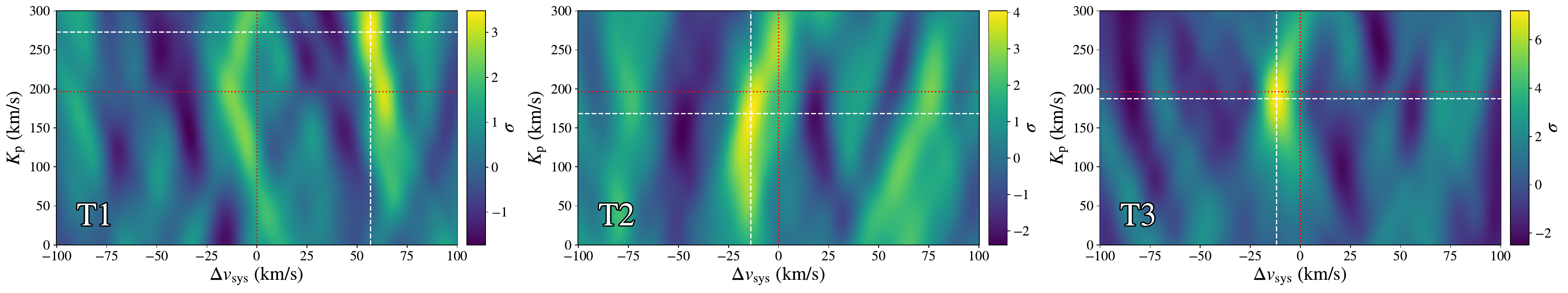}
        \caption{$K_{\rm p}-\Delta v_{\text{sys}}$ maps of VO, for T1, T2, and T3, confirming the novel detection of VO by \citetalias{Pelletier_2023}. The VO atmospheric models were calculated using the optimum parameters from our various atmospheric retrievals and broadened with the respective optimum broadening parameters, using the broadening kernel given by Eq. \ref{eqn:kernel}. The red dotted lines shows the expected planetary velocity values, whereas the white dashed lines shows the maximum $K_{\rm p}$ and $\Delta v_{\text{sys}}$ values.}
        \label{VO_comp}
\end{figure*}
\vspace{3cm}
\begin{figure*}[!htbp]
    \centering
    \includegraphics[width=\textwidth]{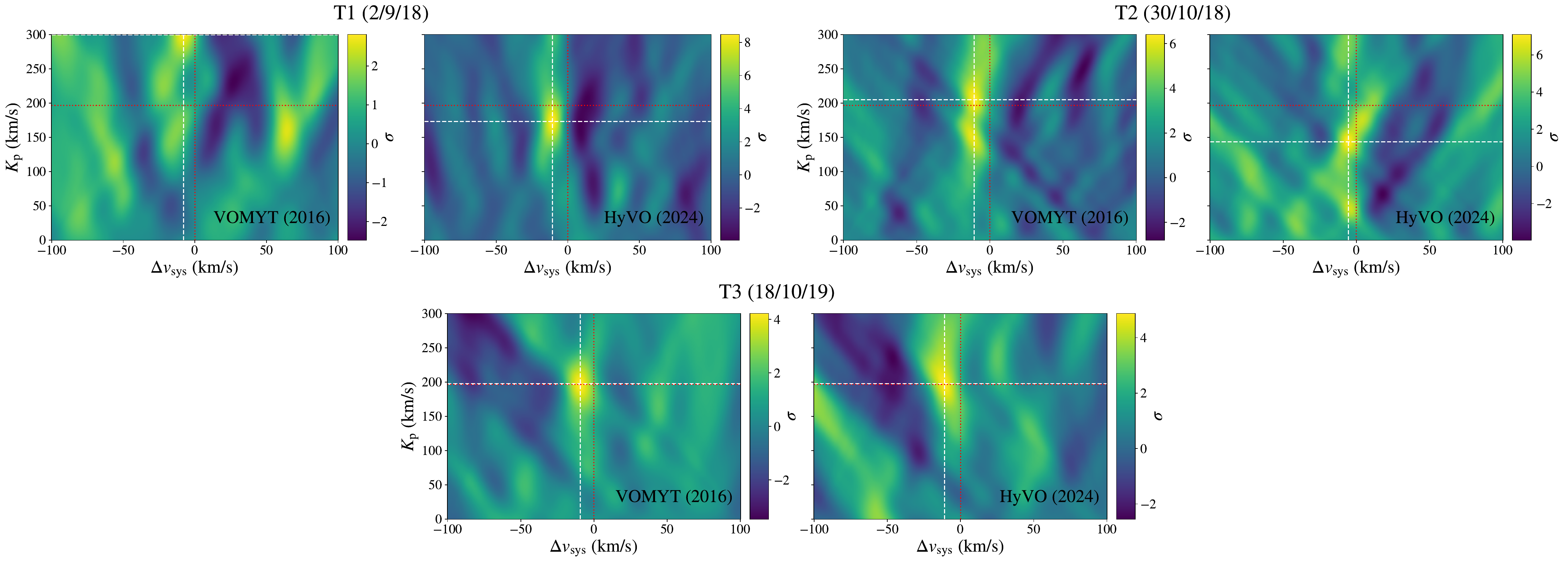}
        \caption{$K_{\rm p}-\Delta v_{\text{sys}}$ maps of VO, for T1, T2, and T3, using both the VOMYT \citep{McKemmish_2016} and HyVO \citep{Bowesman_2024} line lists. The VO cross-sections were calculated at $T = 2800$ K and $P = 0.1$ mbar. The transmission spectra were computed using an isothermal $T$-$P$ profile, and convolved with a 1D Gaussian kernel. The red dotted lines shows the expected planetary velocity values, whereas the white dashed lines shows the maximum $K_{\rm p}$ and $\Delta v_{\text{sys}}$ values.}
        \label{HyVO_VOMYT}
\end{figure*}
\begin{figure*}[!htbp]
    \centering
    \includegraphics[width=\textwidth]{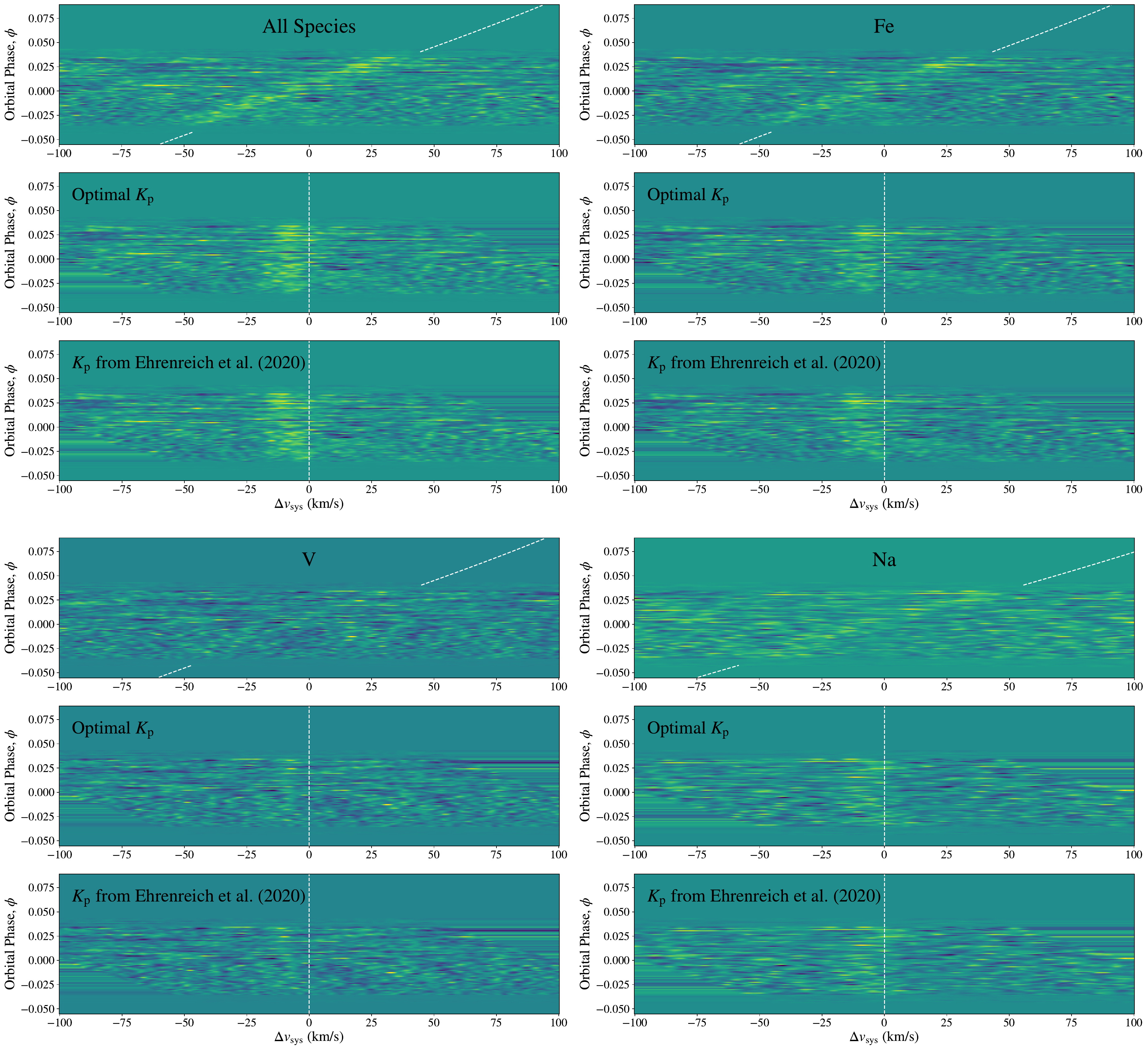}
        \caption{The CC trails of various species for T3, shifted to the optimum $K_{\rm p}$ values, as well as to the $K_{\rm p}$ value from \cite{Ehrenreich_2020}, in order to visualise any apparent ``kink''. The atmospheric models were calculated using the optimum parameters from our various atmospheric retrievals, however, were unbroadened, in order to aid visualisation of any deviation from the planet's Keplerian velocity curve. For clarity, we only plot species in which there was a clearly visible CC trail.}
        \label{CC_trails}
\end{figure*}
\label{appendix_d}
\end{appendix}

\end{document}